\documentclass[prl, twocolumn, superscriptaddress]{revtex4-1}
\usepackage{bm, amsmath, amsfonts, amssymb, braket}
\usepackage{times}
\usepackage{multirow}
\usepackage[dvipdfmx]{graphicx}
\usepackage{float}
\usepackage{color}
\usepackage{comment}

\begin{document}

\newcommand{\ii}{\text{i}}
\newcommand{\hop}{J}

\title{Nonunitary Scaling Theory of Non-Hermitian Localization}

\author{Kohei Kawabata}
	\email{kawabata@cat.phys.s.u-tokyo.ac.jp}
	\affiliation{Department of Physics, University of Tokyo, 7-3-1 Hongo, Bunkyo-ku, Tokyo 113-0033, Japan}

\author{Shinsei Ryu}
  \email{shinseir@princeton.edu}
  \affiliation{Department of Physics, Princeton University, Princeton, New Jersey, 08540, USA}

\date{\today}

\begin{abstract}
Non-Hermiticity can destroy Anderson localization and lead to delocalization even in one dimension. However, the unified understanding of the non-Hermitian delocalization has yet to be established. Here, we develop a scaling theory of localization in non-Hermitian systems. We reveal that non-Hermiticity introduces a new scale and breaks down the one-parameter scaling, which is the central assumption of the conventional scaling theory of localization. Instead, we identify the origin of the unconventional non-Hermitian delocalization as the two-parameter scaling. Furthermore, we establish the threefold universality of non-Hermitian localization based on reciprocity; reciprocity forbids the delocalization without internal degrees of freedom, whereas symplectic reciprocity results in a new type of symmetry-protected delocalization.
\end{abstract}

\maketitle
Anderson localization~\cite{Anderson-58} is the disorder-induced localization of coherent waves and plays an important role in transport phenomena of
condensed matter~\cite{Lee-review, Evers-review},
light~\cite{Segev-review}, and cold atoms~\cite{Billy-08, Roati-08}.
A unified understanding of Anderson localization is provided by the scaling theory~\cite{Thouless-74, Abrahams-79, *Anderson-80}.
On the basis of the one-parameter-scaling hypothesis of the conductance
with respect to to the system size, it describes the criticality of localization transitions in three dimensions
and predicts the absence of delocalization in one and two dimensions.
Symmetry further changes the universality class
of localization.
For example, time-reversal symmetry (reciprocity) in the presence of spin-orbit interaction enables delocalization even in two dimensions~\cite{Hikami-80}; chiral (sublattice) symmetry enables delocalization of zero modes even in one dimension~\cite{Dyson-53, AZ-97, Balents-97, Brouwer-98, *Brouwer-00, Ryu-04}.

Meanwhile, the physics of non-Hermitian systems has attracted
considerable interest in recent years~\cite{Konotop-review, Christodoulides-review, Ota-review, Bergholtz-review}.
Non-Hermiticity originates from exchanges of energy or particles
with an environment and leads to rich properties unique to
particle-number-nonconserving 
systems in dynamics~\cite{Makris-08, Klaiman-08, Guo-09,
  Ruter-10, Lin-11, Regensburger-12, Feng-13, Peng-14, Wiersig-14,
  Peng-16, Ashida-17, Hodaei-17, Chen-17, Kawabata-17, *Xiao-19, Rivet-18,
  Nakagawa-18, Dora-19, Li-19}
and topology~\cite{Rudner-09, *Zeuner-15, Sato-12, *Esaki-11, Hu-11,
  Schomerus-13, *Poli-15, Zhen-15, *Zhou-18, Weimann-17, Lee-16, Leykam-17,
  Xu-17, Xiao-17, St-Jean-17, Shen-18, *Kozii-17, Bahari-17, Harari-18,
  *Bandres-18, MartinezAlvarez-18, Gong-18, *Kawabata-19, Yao-18-SSH, *Yao-18-Chern, Kunst-18,
  Lee-19, Kunst-19, KSUS-19, *KBS-19, ZL-19, Borgnia-20, Yokomizo-19, Okuma-19,
  Zhao-19, Zhang-19, OKSS-20}.
Anderson localization was also investigated in non-Hermitian systems
with asymmetric hopping~\cite{Hatano-Nelson-96, *Hatano-Nelson-97, *Hatano-Nelson-98, Efetov-97,
  Feinberg-97, *Feinberg-99, Brouwer-97, Goldsheid-98, Nelson-98, *Amir-16,
  Mudry-98, Fukui-98, Yurkevich-99, Longhi-15, McDonald-18,
  Hamazaki-19, Zeng-20, Jiang-20}
and gain or loss~\cite{Freilikher-94,
  Beenakker-96, *Paasschens-96, Bruce-96, Longhi-19L, Tzortzakakis-20, Huang-20},
the latter of which is directly relevant to random lasers~\cite{Wiersma-review-NatPhys08}.
Even in the presence of non-Hermiticity, random lasers in one dimension never exhibit delocalization similarly to the Hermitian case.
By contrast, a non-Hermitian extension of the Anderson model with asymmetric hopping, which was first investigated by Hatano and Nelson~\cite{Hatano-Nelson-96, *Hatano-Nelson-97, *Hatano-Nelson-98}, exhibits delocalization in one dimension.
Importantly, this implies the breakdown of the conventional scaling theory of localization, which predicts the absence of delocalization in one dimension.
In fact, since Anderson localization results from the destructive interference of coherent waves, non-Hermiticity should lead to decoherence and destroy Anderson localization. 
However, it remains unclear how non-Hermiticity changes
the scaling theory of localization, and a unified understanding
of non-Hermitian localization has yet to be obtained.

\begin{figure}[t]
\centering
\includegraphics[width=80mm]{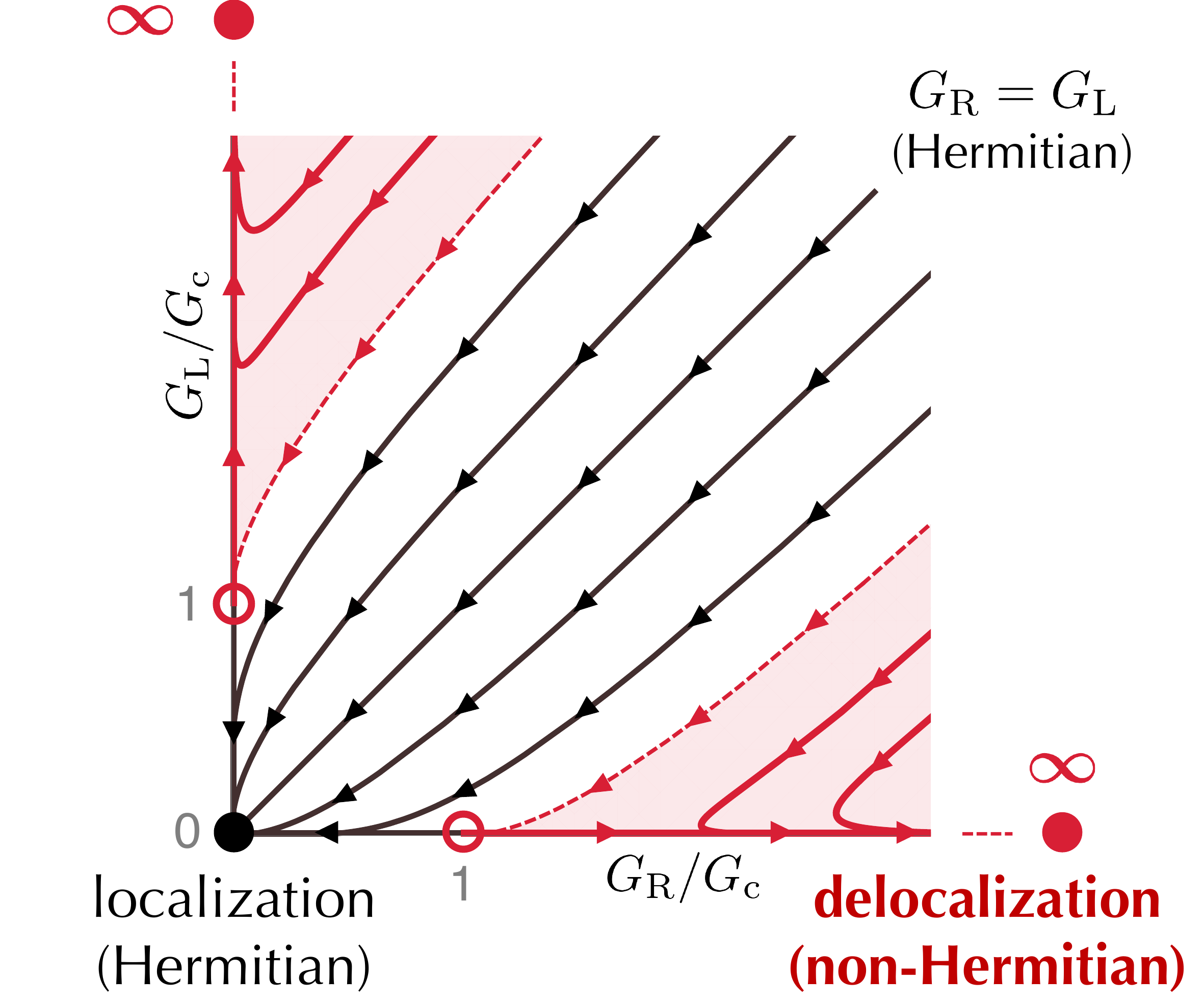} 
\caption{Two-parameter scaling of non-Hermitian localization. The renormalization-group flow is shown according to the conductance $G_{\rm R}$ from the left to the right and the conductance $G_{\rm L}$ from the right to the left. The system size $L$ increases along with the arrows. While localization with $\left( G_{\rm R}, G_{\rm L} \right) = \left( 0, 0 \right)$ (black dot) occurs in Hermitian systems ($G_{\rm R} = G_{\rm L}$), delocalization with $\left( G_{\rm R}, G_{\rm L} \right) = \left( \infty, 0 \right), \left( 0, \infty \right)$ (red dots) arises for sufficiently strong non-Hermiticity.}
	\label{fig: RG}
\end{figure}

In this Letter, we develop a scaling theory of localization in non-Hermitian
systems. On the basis of the random-matrix approach for nonunitary scattering
matrices, we reveal that non-Hermiticity introduces a new scale
and breaks down the one-parameter-scaling hypothesis. Instead, we
demonstrate the two-parameter scaling
(Fig.~\ref{fig: RG}),
which is the origin of the unconventional non-Hermitian delocalization.
Furthermore, we establish the threefold universality of non-Hermitian
localization according to reciprocity
(Table~\ref{tab: reciprocity}).
While non-Hermitian systems exhibit unidirectional delocalization in the absence of symmetry, reciprocity forbids it without internal degrees of freedom, which explains the absence of delocalization in random lasers. We also find a new universality class of localization transitions: bidirectional delocalization protected by symplectic reciprocity.

\paragraph{Non-Hermitian delocalization.\,---}

In the conventional scaling theory of localization~\cite{Abrahams-79}, we consider the dependence of the conductance $G$ on the length scale $L$.
A sufficiently small system is diffusive and described by Ohm's law (Boltzmann equation), leading to $G \propto L^{d-2}$ in $d$ dimensions.
For a sufficiently large system, on the other hand, the wave coherence is relevant and Anderson localization can occur, leading to $G \propto e^{-\alpha L}$ ($\alpha > 0$). The transition between these two regimes can be understood by the scaling function $\beta \left( G \right) := d\log G/d\log L$. In the localized regime, it is given as $\beta \left( G \right) = \log G < 0$ and hence the conductance $G$ gets smaller with increasing the system length $L$. We have $\beta \left( G \right) = d-2$ in the diffusive regime, which is positive (negative) for $d>2$ ($d<2$). Consequently, a localization transition occurs in three dimensions
at $G=G_c$ where $\beta \left( G_{\rm c} \right) = 0$;
by contrast, no transitions occur in one dimension since $\beta \left( G \right)$ is always negative and $G$ monotonically decreases in both diffusive and localized regimes.

Non-Hermiticity gives rise to
a new regime that has no analogs in
particle-number-conserving systems.
In fact, it describes coupling to an external environment and can lead to
amplification (lasing), resulting in $G \propto e^{\gamma L}$ with the
amplification rate $\gamma > 0$.
In such a regime, we have $\beta \left( G \right) = \log G > 0$ in arbitrary
dimensions, and hence delocalization is possible even in one dimension.
The amplifying regime can 
arise from nonunitarity of scattering matrices.
In Hermitian systems, unitarity is imposed on scattering matrices as a direct
consequence
of conservation of particle numbers, and the transmission amplitudes
cannot exceed one. 
In non-Hermitian systems, by contrast,
such a constraint is absent and the conductance $G$ can be arbitrarily large, which enables the amplification as $G \propto e^{\gamma L}$.

The delocalization in the amplifying regime can also be understood by the Thouless criterion~\cite{Thouless-74, RH-pc}. In the diffusive regime, it takes the Thouless time $t_{\rm Th} \propto L^{2}$ for a particle to reach one end from the other in a system of size $L^{d}$. To realize this diffusive transport, $t_{\rm Th}$ should be smaller than the time scale $\Delta t \propto \left( \Delta E \right)^{-1}$ determined by the level spacing $\Delta E \propto L^{-d}$ of the spectrum. Because of $t_{\rm Th}/\Delta t \propto L^{2-d}$, this is possible in three dimensions but impossible in one dimension. In the amplifying regime, on the other hand, particle inflow from the environment enables the ballistic transport,
and the relevant time scale is $t_{\rm N} \propto L$.
Because of $t_{\rm N}/\Delta t \propto L^{1-d}$,
$t_{\rm N}$ is comparable to $\Delta t$ even in one dimension,
which results in the delocalization. Saliently, an additional relevant scale accompanies the amplifying regime, which implies the breakdown of the one-parameter-scaling hypothesis~\cite{Thouless-74, Abrahams-79}, as discussed below.

\paragraph{Scaling equations.\,---}
To uncover universal behavior of Anderson localization in non-Hermitian systems, we revisit the Hatano-Nelson model~\cite{Hatano-Nelson-96, *Hatano-Nelson-97, *Hatano-Nelson-98} and derive the scaling equations for transport properties. We show that the scaling behavior should be understood in terms of two parameters rather than one parameter. 
On the basis of this understanding, we later discuss Anderson localization for other symmetry classes and find new universality classes.
Our scaling theory also explains the different universality classes between the Hatano-Nelson model and random lasers.

The Hatano-Nelson model~\cite{Hatano-Nelson-96, *Hatano-Nelson-97, *Hatano-Nelson-98} reads
\begin{equation}
\hat{H} = \sum_{n} \left\{ - \frac{1}{2} \left( \hat{c}_{n+1}^{\dag} \hop_{\rm R} \hat{c}_{n} + \hat{c}_{n}^{\dag} \hop_{\rm L} \hat{c}_{n+1} \right) + \hat{c}_{n}^{\dag} M_{n} \hat{c}_{n} \right\},
	\label{eq: Hamiltonian - lattice}
\end{equation}
where $\hat{c}_{n}$ ($\hat{c}_{n}^{\dag}$) annihilates (creates) a fermion at site $n$, $J_{\rm R} := J+\gamma/2$
($J_{\rm L} := J-\gamma/2$) describes the hopping from the left to the right (from the right to the left), and $M_{n} \in \mathbb{R}$ is the random potential
at
site $n$.
The asymmetry $\gamma$ of the hopping can be introduced, for example, in open photonic systems~\cite{Chen-17, Longhi-15} and cold atoms with dissipation~\cite{Gong-18}. Whereas eigenstates are localized for weak $\gamma$, they can be delocalized for strong $\gamma$. In the literature, the complex spectrum~\cite{Hatano-Nelson-96, *Hatano-Nelson-97, *Hatano-Nelson-98, Feinberg-97, *Feinberg-99, Brouwer-97, Goldsheid-98, Hamazaki-19},
the conductance~\cite{Brouwer-97, Yurkevich-99}, and the chiral transport~\cite{Longhi-15, McDonald-18} were investigated for this lattice model. Nevertheless, the scaling theory has not been fully formulated.

The nature of the non-Hermitian delocalization should not depend on 
specific details of the model but solely on symmetry. To understand such a
universal feature, we construct a continuum model from the Hatano-Nelson model.
To this end, we focus on the narrow shell around the band center $\mathrm{Re}\,E
= 0$ and decompose the fermions by $\hat{c}_{n} = e^{\ii k_{\rm F} n}
\hat{\psi}_{\rm R} + e^{-\ii k_{\rm F} n} \hat{\psi}_{\rm L}$ ($k_{\rm F} = \pi/2$). Here, $\hat{\psi}_{\rm R}$ and
$\hat{\psi}_{\rm L}$ are the right-moving and left-moving fermions on
the two Fermi points (valleys), respectively.
Assuming that $\hat{\psi}_{\rm R}$ and $\hat{\psi}_{\rm L}$ vary slowly in
space, we have the continuum model
$\hat{H} = \int dx\, ( \hat{\psi}_{\rm R}^{\dag}~\hat{\psi}_{\rm L}^{\dag} )\,h_{\rm A} (
\hat{\psi}_{\rm R}~\hat{\psi}_{\rm L} )^{T}$
with
\begin{equation}
h_{\rm A} = \left( -\ii \partial_{x} + \ii \gamma/2 \right) \tau_{3} + m_{0} \left( x \right) + m_{1} \left( x \right) \tau_{1},
	\label{eq: continuum - A}
\end{equation}
where Pauli matrices $\tau_{i}$'s 
describe the two valley degrees of freedom. 
We assume that $m_{0}$ and $m_{1}$ are Gaussian disorder
that satisfies $\braket{m_{i} \left( x \right)} = 0$ and $\braket{m_{i} \left( x \right) m_{j} \left( x' \right)} = 2\mu_{i} \delta_{ij} \delta \left( x-x' \right)$
with $\mu_{i} > 0$ and the ensemble average $\braket{\star}$.
Although we begin with the Hatano-Nelson model,
we emphasize that $h_{\rm A}$ does not depend on its specific details but
universally on symmetry.
Generic non-Hermitian systems without symmetry including
$h_{{\rm A}}$ are defined to belong to class A in the 38-fold classification of internal symmetry~\cite{BL-02, KSUS-19, HN-TRS}. 

Now, we formulate the scaling equations (functional renormalization group equations).
The conductance $G_{\rm R}$ from the left to the right
($G_{\rm L}$ from the right to the left)
is given by the corresponding transmission eigenvalue $T_{\rm R}$ ($T_{\rm L}$) 
according to the Landauer formula~\cite{Datta-textbook}.
Then, we consider the incremental changes of $T_{\rm R/L}$,
in addition to the reflection eigenvalue
$R_{\rm L}$ from the left to the left ($R_{\rm R}$ from the right to the right), upon attachment
of a thin slice in which the scattering can be treated perturbatively. 
Such attachment renormalizes the probability distribution of $T_{\rm R/L}$ and $R_{\rm L/R}$, resulting in its scaling (Fokker-Planck) equation according to the system size $L$~\cite{supplement}.
It provides all the information about the transmission eigenvalues $T_{\rm R/L}$ and the conductances $G_{\rm R/L}$. 
In the Hermitian case, the scaling equations were obtained by Dorokhov, and by Mello, Pereyra, and Kumar~\cite{Dorokhov-82, Mello-88, Beenakker-review, *Beenakker-review-sc}.

For the continuum model $h_{\rm A}$, we find that non-Hermiticity $\gamma$ amplifies one of $T_{\rm R}$ and $T_{\rm L}$ and attenuates the other, but does not have significant influence on their phases. As a result, we have~\cite{supplement}
\begin{equation}
\frac{\braket{dT_{\rm R/L}}}{dL} = \pm \gamma T_{\rm R/L} - \frac{T_{\rm R/L} \left( 1-R_{\rm L/R} \right)}{\ell},
	\label{eq: scaling}
\end{equation}
where $\ell := 1/2\mu_{1}$ is the mean free path determined by the disorder strength. 
The ensemble average $\braket{\star}$ is taken over the attached thin slice and the phases of the scattered waves for given $T_{\rm R/L}$ and $R_{\rm L/R}$.
This scaling equation~(\ref{eq: scaling}) implies that the transmission amplitudes are given
as $T_{\rm R/L} = e^{\pm \gamma L} \tilde{T}$ with the transmission amplitude
$\tilde{T}$ in Hermitian systems.
For $L \gg \ell$, the conductance fluctuations become as large as the averages $\braket{G}$, which no longer represent the conductances of a single sample. In fact, the conductance distributions are broad and asymmetric, and follow the log-normal distributions. Consequently, the typical conductances are $G^{\rm typ} := e^{\braket{\log G}}$ instead of $\braket{G}$.
Because of $\tilde{G}^{\rm typ}/G_{\rm c} \sim e^{-L/\ell}$ for $L \gg \ell$ in the Hermitian case~\cite{Dorokhov-82, Mello-88, Beenakker-review, *Beenakker-review-sc}, the typical conductances in the non-Hermitian case are $G_{\rm R/L}^{\rm typ}/G_{\rm c} \sim e^{\left( \pm \gamma -
    1/\ell \right) L}$.
Thus, either one of the two conductances exhibits delocalization.
For $\gamma \geq 0$, for example, 
$G_{\rm R}^{\rm typ}$ diverges 
for $L \to \infty$ above the transition point $\gamma = \gamma_{\rm c} := 1/\ell$,
around which the critical behavior $| G_{\rm R}^{\rm typ} - G_{\rm c} |/G_{\rm c} \propto \left| \gamma - \gamma_{\rm c} \right|$ appears.

\paragraph{Two-parameter scaling.\,---}

In Hermitian systems, the scaling equations and the conductance $\tilde{G}$
depend solely on $L/\ell$.
This confirms the one-parameter-scaling hypothesis, which underlies the absence of delocalization in one dimension~\cite{Thouless-74, Abrahams-79}.
However, the obtained scaling equation~(\ref{eq: scaling}) clearly indicates the emergence of the additional scale $\gamma$ due to non-Hermiticity.
In fact, non-Hermiticity leads to the distinction between $G_{\rm R}$ and $G_{\rm L}$, which is impossible in Hermitian systems by conservation of particle numbers. From Eq.~(\ref{eq: scaling}), we show in Fig.~\ref{fig: RG} the renormalization-group flow based on both $G_{\rm R}$ and $G_{\rm L}$. 
In addition to the fixed point $\left( G_{\rm R}, G_{\rm L} \right) = \left( 0, 0 \right)$ for the localized phase, a pair of additional fixed points $\left( G_{\rm R}, G_{\rm L} \right) = \left( G_{\rm c}, 0 \right), \left( 0, G_{\rm c} \right)$ emerges away from $G_{\rm R} = G_{\rm L}$.
As a result, delocalization with $\left( G_{\rm R}, G_{\rm L} \right) = \left( \infty, 0 \right), \left( 0, \infty \right)$ is possible for sufficiently strong non-Hermiticity.
Therefore, the emergence of the new scale and the breakdown of the one-parameter scaling are the origin of the non-Hermitian delocalization in one dimension. 

It is also notable that the average conductances are $\braket{G_{\rm R/L}}/G_{\rm c} \sim e^{\left( \pm \gamma - 1/4\ell \right) L}$ since the Hermitian counterpart is $\braket{\tilde{G}}/G_{\rm c} \sim e^{- L/4\ell}$~\cite{Dorokhov-82, Mello-88, Beenakker-review, *Beenakker-review-sc}. Hence, $\braket{G_{\rm R}}$ exhibits critical behavior at $\gamma = 1/4\ell$, which is different from the critical point $\gamma = 1/\ell$ of the typical conductance $G_{\rm R}^{\rm typ}$~\cite{Yurkevich-99}.
Such a difference of the critical points is
another manifestation of the breakdown of the one-parameter scaling.
In fact, if the scaling equations are described solely by a single parameter $\ell$, both $\braket{G_{\rm R}}$ and $G_{\rm R}^{\rm typ}$ are functions of $L/\ell$, and hence their critical points should coincide with each other. The different critical points of $\braket{G_{\rm R}}$ and $G_{\rm R}^{\rm typ}$ imply the two different length scales $\ell$ and $\gamma^{-1}$.

\paragraph{Threefold universality by reciprocity.\,---}

Symmetry can further change
the universality class of Anderson localization.
In particular, reciprocity, defined by
$\mathcal{T} H^{T} \mathcal{T}^{-1} = H$ with a unitary matrix $\mathcal{T}$, is
fundamental symmetry relevant to localization.
For example, reciprocity with $\mathcal{T}\mathcal{T}^{*} = +1$ ($-1$)
enhances (suppresses) 
localization and shortens (lengthens) localization lengths in Hermitian wires in
quasi-one dimension~\cite{Dorokhov-82, Mello-88, Beenakker-review,
  *Beenakker-review-sc}.
Moreover, symplectic reciprocity with $\mathcal{T}\mathcal{T}^{*} = -1$ enables delocalization even in two dimensions~\cite{Hikami-80}, although delocalization is forbidden without symmetry protection. Here, we uncover the threefold universality of non-Hermitian localization based on reciprocity (Table~\ref{tab: reciprocity}). 
As demonstrated below, the influence of reciprocity is more dramatic than the Hermitian case.

\begin{table}[t]
	\centering
	\caption{Threefold universality of non-Hermitian localization based on reciprocity. The types of delocalization and the typical conductances for $L \gg \ell$ are shown according to non-Hermiticity $\gamma$ and the mean free path $\ell > 0$.}
     \begin{tabular}{cccc} \hline \hline
     ~~Class~~ & ~~Symmetry~~ & ~~Delocalization~~ & ~~Conductances~~  \\ \hline
     A & No & Unidirectional & $e^{( \pm \gamma - 1/\ell ) L}$ \\
     $\text{AI}^{\dag}$ & $H^{T} = H$ & No & $e^{-L/\ell}$ \\
     $\text{AII}^{\dag}$ & ~~$\sigma_{2} H^{T} \sigma_{2}^{-1} = H$~~ & Bidirectional & $e^{( \left| \gamma \right| - 1/\ell ) L}$ \\ \hline \hline
    \end{tabular}
	\label{tab: reciprocity}
\end{table}

We consider a non-Hermitian continuum model
\begin{equation}
h_{\text{AI}^{\dag}} = -\ii \tau_{3} \partial_{x} + m_{0} \left( x \right) + \left( m_{1} \left( x \right) + \ii \gamma/2 \right) \tau_{1},
	\label{eq: continuum - A1}
\end{equation}
which respects $\tau_{1} h_{\text{AI}^{\dag}}^{T} \tau_{1}^{-1} = h$ and hence
belongs to class $\text{AI}^{\dag}$ (orthogonal class)~\cite{KSUS-19, supplement}.
Notably, the asymmetry between the valleys [i.e., $\ii \left( \gamma/2 \right) \tau_{3}$ term in Eq.~(\ref{eq: continuum - A})]
is forbidden because of reciprocity, which leads to $G_{\rm R} = G_{\rm L}$ even in non-Hermitian systems. Thus, the nonunitary fixed points away from $G_{\rm R} = G_{\rm L}$ in Fig.~\ref{fig: RG} cannot be reached, and the unidirectional delocalization is forbidden.
In terms of the scaling equations, reciprocity-preserving non-Hermiticity is irrelevant by the ensemble average over disorder, whereas reciprocity-breaking non-Hermiticity gives rise to an additional scale.
Consequently, the universality in class $\text{AI}^{\dag}$ is the same as the Hermitian counterpart, which contrasts with class A.
The continuum model in Eq.~(\ref{eq: continuum - A1}) describes disordered wires with gain or loss (i.e., complex onsite potential), including random lasers~\cite{Wiersma-review-NatPhys08}. Reciprocity underlies the absence of delocalization in random lasers.

On the other hand, reciprocal systems with $\mathcal{T}\mathcal{T}^{*} = -1$
instead of $\mathcal{T}\mathcal{T}^{*} = +1$ are defined to belong to class
$\text{AII}^{\dag}$ (symplectic class)~\cite{KSUS-19}. Although reciprocity
imposes $G_{\rm R} = G_{\rm L}$ also in this case, an important distinction in the 
symplectic class is Kramers degeneracy,
which gives rise to 
a new type of non-Hermitian delocalization protected by reciprocity. The corresponding continuum model is 
\begin{equation}
h_{\text{AII}^{\dag}} = \left( -\ii \partial_{x} + \Delta \sigma_{1} + \ii \left( \gamma /2 \right) \sigma_{3} \right) \tau_{3} + m_{0} \left( x \right) + m_{1} \left( x \right) \tau_{1},
	\label{eq: continuum - A2}
\end{equation}
which respects $\left( \sigma_{2} \tau_{1} \right) h_{\text{AII}^{\dag}}^{T}
\left( \sigma_{2} \tau_{1} \right)^{-1} = h_{\text{AII}^{\dag}}$. Here, Pauli
matrices $\sigma_{i}$'s describe the internal degrees of freedom such as spin.
The scaling equations can be obtained in a similar manner to class
A~\cite{supplement}.
In this case, one of the Kramers pair is amplified to
the right while the other to the left because of non-Hermiticity. We then have $G_{\rm R}^{\rm typ}/G_{\rm c} =
G_{\rm L}^{\rm typ}/G_{\rm c} \sim ( e^{\gamma L} + e^{-\gamma L} )\,e^{-L/\ell}
\sim e^{\left( \left|\gamma\right| - 1/\ell \right) L}$ for $L \gg \ell$. Thus,
the eigenstates are bidirectionally delocalized in contrast to
classes A and $\text{AI}^{\dag}$. 
Without symmetry, one of the transmitted channels dominates the other, and non-Hermitian delocalization is unidirectional. Hence, the bidirectional delocalization arises only in the presence of symplectic reciprocity.
Despite $G_{\rm R} = G_{\rm L}$, the conductance of one channel serves as $G_{\rm R}$ and that of the corresponding Kramers partner serves as $G_{\rm L}$ in the two-parameter scaling shown in Fig.~\ref{fig: RG}, the sum of which yields the total conductance.

While reciprocity is equivalent to time-reversal symmetry $\mathcal{T} H^{*} \mathcal{T}^{-1} = H$ in Hermitian systems, this is not the case in non-Hermitian systems.
The corresponding symmetry classes with time-reversal symmetry 
are classes AI and AII~\cite{KSUS-19}.
The universality of non-Hermitian localization is also different
depending on whether one imposes 
time-reversal symmetry or reciprocity. In fact, time-reversal symmetry does not change the universality of the non-Hermitian localization~\cite{supplement}, whereas reciprocity can forbid or enhance it as discussed above.
Reciprocity also leads to threefold universality of non-Hermitian random matrices~\cite{HKKU-19}.

\paragraph{Lattice models.\,---}

\begin{figure}[t]
\centering
\includegraphics[width=86mm]{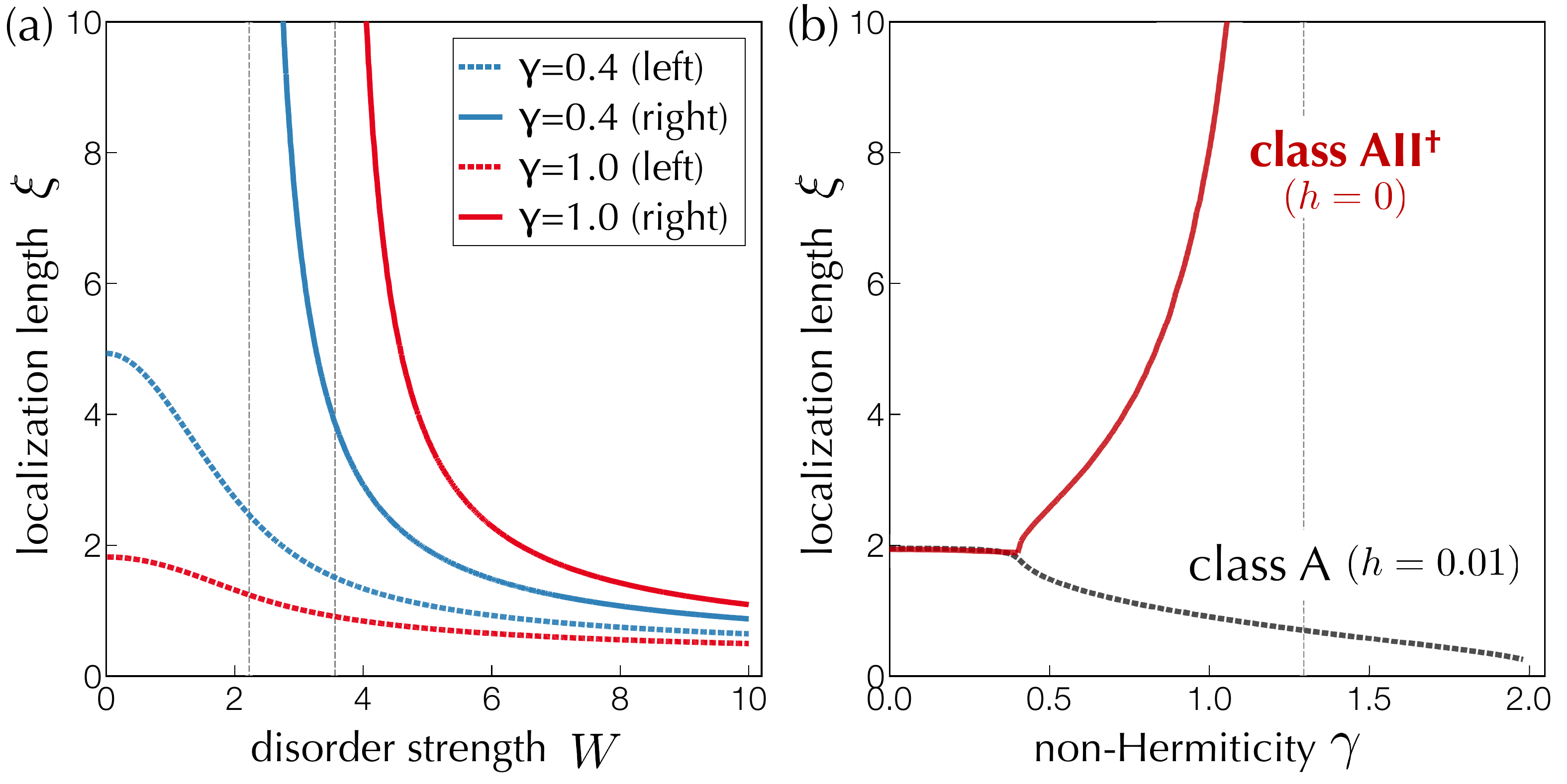} 
\caption{Non-Hermitian localization on lattices ($L=5000$,
  $\hop = 1.0$, $E=0$).
  The disordered onsite potential is uniformly distributed over $\left[ -W/2, W/2 \right]$, and each datum is averaged over $1000$ samples. (a)~Hatano-Nelson model (class A). For $\gamma > 0$, the right localization length diverges at a transition point, whereas the left localization length remains finite. The transition points are $W_{\rm c} = 2.22$ ($\gamma=0.4$) and $W_{\rm c} = 3.56$ ($\gamma=1.0$). (b)~Symplectic Hatano-Nelson model (class $\text{AII}^{\dag}$; $\Delta = 0.2$, $W = 4.0$). Both right and left localization lengths diverge at the transition point $\gamma_{\rm c} = 1.30$ (red solid curve). The delocalization vanishes in the presence of a reciprocity-breaking perturbation $h=0.01$ (black dotted curve).}
	\label{fig: localization-length}
\end{figure}

To confirm our nonunitary scaling theory, we numerically investigate non-Hermitian lattice models by the transfer-matrix
method~\cite{Ohtsuki-review, supplement}. In general, a wave function localized
around site $n=n_{0}$ is proportional to $e^{- \left| n - n_{0}
  \right|/\xi_{\rm L} (\xi_{\rm R})}$ for $n < n_{0}$ ($n > n_{0}$).
While the two localization lengths $\xi_{\rm L}$ and $\xi_{\rm R}$ are equivalent in Hermitian systems, they are different
in a similar manner to
the conductances $G_{\rm L}$ and $G_{\rm R}$.
Figure~\ref{fig: localization-length}\,(a) shows the localization lengths for the Hatano-Nelson model in Eq.~(\ref{eq: Hamiltonian - lattice}).
For $\gamma \geq 0$, the right localization length $\xi_{\rm R}$ diverges at a critical point, whereas the left localization length $\xi_{\rm L}$ remains finite, which is a signature of the unidirectional delocalization. 
Around the critical point, $\xi_{\rm R}$ diverges as $\xi_{\rm R} \propto \left| \gamma - \gamma_{\rm c} \right|^{-1}$.

A symplectic extension of the Hatano-Nelson model is given by Eq.~(\ref{eq: Hamiltonian - lattice})
with $\hop_{\rm R} := \hop - \ii \Delta \sigma_{1} + \gamma \sigma_{3}/2$, $\hop_{\rm L} := \hop + \ii \Delta \sigma_{1} - \gamma \sigma_{3}/2$, and $M_{n} := m_{n} + h\sigma_{3}$.
This lattice model with $h=0$ corresponds to the continuum model in Eq.~(\ref{eq: continuum - A2}). In contrast to the original Hatano-Nelson model, we have $\xi_{\rm L} = \xi_{\rm R}$ for $h=0$ because of reciprocity. As a result, both $\xi_{\rm L}$ and $\xi_{\rm R}$ diverge at a critical point [Fig.~\ref{fig: localization-length}\,(b)], which is a signature of the bidirectional delocalization. Because of the reciprocity-protected nature, 
the delocalization vanishes even in the presence of a small reciprocity-breaking perturbation $h \neq 0$, which is unique to the symplectic class.

\paragraph{Chiral symmetry and sublattice symmetry.\,---}

In the presence of chiral or sublattice symmetry, zero modes can be delocalized even in Hermitian systems in one dimension, accompanied by Dyson's singularity~\cite{Dyson-53}.
Similarly to time-reversal symmetry and reciprocity, chiral symmetry and sublattice symmetry are distinct from each other in non-Hermitian systems, the former (latter) of which corresponds to class AIII ($\text{AIII}^{\dag}$)~\cite{KSUS-19}. For example, a random hopping model with gain or loss respects chiral symmetry, while a random asymmetric hopping model respects sublattice symmetry. In the presence of chiral symmetry $\tau_{1} h^{\dag}_{\text{AIII}} \tau_{1}^{-1} = - h_{\text{AIII}}$, non-Hermiticity is found not to change the universality of the delocalization~\cite{supplement}; by contrast, in the presence of sublattice symmetry $\tau_{1} h_{\text{AIII}^{\dag}} \tau_{1}^{-1} = - h_{\text{AIII}^{\dag}}$, non-Hermiticity
enables the unidirectional delocalization in a similar manner to class A. In fact, the asymmetry between the valleys is allowed 
for sublattice symmetry, but forbidden for chiral symmetry.

\paragraph{Discussions.\,---}

Transport  
phenomena of disordered systems,
including Anderson localization and transitions,
enjoy universality in various scaling limits
that is
governed only by a few physical parameters.
This is embodied by the one-parameter scaling
of localization~\cite{Thouless-74, Abrahams-79}.
In this Letter, we have demonstrated that non-Hermiticity breaks
it down and leads to the
two-parameter scaling, which generally describes the unconventional
non-Hermitian delocalization.
While we limit ourselves to the single-channel case in this Letter,
it is meaningful to consider the limit of thick wires in order to fully uncover
universal properties---we leave this as a future problem. 

In our nonunitary two-parameter scaling, the critical exponents are integers, which contrast with the more complicated exponents
in the two-parameter scaling of the quantum Hall transition~\cite{QHE_review, Huckestein-review, Kramer-review, scaling-dimension}.
On the other hand, these two scaling theories share similarities from a topological perspective.
In particular, 
the Hatano-Nelson model is characterized by a topological invariant unique to non-Hermitian systems~\cite{Gong-18,Kawabata-19,KSUS-19}.
In our continuum model, this topological invariant is
$\mathrm{sgn}\, \gamma$, similar to 
the Hall conductivity given by the Dirac mass term.
An open problem is to formulate an effective field theory
for the nonunitary two-parameter scaling,
akin to the nonlinear sigma model augmented with a topological term for the quantum Hall transition~\cite{QHE_review, Huckestein-review, Kramer-review}.
In this regard, it is worth pointing out that topological field theory
descriptions for non-Hermitian
systems have been recently proposed~\cite{2020arXiv201111449Kmisc}.

\medskip
K.K. thanks Zongping Gong, Hideaki Obuse, Tomi Ohtsuki, Masatoshi Sato, and Masahito Ueda for
helpful discussions. In particular, K.K. appreciates stimulating discussions
with Ryusuke Hamazaki. K.K. is supported by KAKENHI Grant No.~JP19J21927 from
the Japan Society for the Promotion of Science (JSPS).
S.R. is supported by the National Science Foundation under award number DMR-2001181, and by a Simons Investigator Grant from the Simons Foundation (Award Number: 566116).

\bibliography{NH-localization}

\begin{thebibliography}{118}
\bibitem [{HN-()}]{HN-TRS}%
  \BibitemOpen
  \href@noop {} {}\bibinfo {note} {{Despite time-reversal symmetry of $h_{\rm
  A}$ (i.e., $\tau_{1} h_{\rm A}^{*} \tau_{1}^{-1} = h_{\rm A}$), it does not
  change the universality of localization.}}\BibitemShut {Stop}%
\bibitem [{sca()}]{scaling-dimension}%
  \BibitemOpen
  \href@noop {} {}\bibinfo {note} {The simple scaling dimensions appear to be
  due to one dimension. It is worthwhile to study whether non-Hermiticity
  induces nontrivial scaling dimensions in higher dimensions.}\BibitemShut
  {Stop}%
\end{thebibliography}

\widetext
\pagebreak

\renewcommand{\theequation}{S\arabic{equation}}
\renewcommand{\thefigure}{S\arabic{figure}}
\renewcommand{\thetable}{S\arabic{table}}
\setcounter{equation}{0}
\setcounter{figure}{0}
\setcounter{table}{0}

\begin{center}
{\bf \large Supplemental Material for 
``Nonunitary Scaling Theory of Non-Hermitian Localization"}
\end{center}

\section{SI.~Non-Hermitian Hamiltonians and nonunitary scattering matrices}

We consider a non-Hermitian disordered system $H$ in one dimension
connected
to
two ideal leads.
A wave incident on the disordered region from
the
left (right) is 
\begin{equation}
a_{\rm in}^{+} := \left( a_{1}^{+}~a_{2}^{+}~\cdots~a_{N}^{+} \right)^{T}
\quad 
\left[ b_{\rm in}^{-} := \left( b_{1}^{-}~b_{2}^{-}~\cdots~b_{N}^{-} \right)^{T} \right],
\end{equation}
and the reflected and transmitted waves scattered to
the
right (left) are 
\begin{equation}
b_{\rm out}^{+} := \left( b_{1}^{+}~b_{2}^{+}~\cdots~b_{N}^{+} \right)^{T}
\quad 
\left[ a_{\rm out}^{-} := \left( a_{1}^{-}~a_{2}^{-}~\cdots~a_{N}^{-} \right)^{T} \right].
\end{equation}
The scattering matrix $S$ relates these incident and scattered waves by
\begin{equation}
\left( \begin{array}{@{\,}c@{\,}} 
	a_{\rm out}^{-} \\ b_{\rm out}^{+} \\ 
	\end{array} \right) 
= S \left( \begin{array}{@{\,}c@{\,}} 
	a_{\rm in}^{+} \\ b_{\rm in}^{-} \\ 
	\end{array} \right),\quad
S := \left( \begin{array}{@{\,}cc@{\,}} 
	r_{\rm L} & t_{\rm L} \\
	t_{\rm R} & r_{\rm R} \\ 
	\end{array} \right),
\end{equation}
where $r_{\rm L}$ ($r_{\rm R}$) is an $N \times N$ invertible matrix that
describes the reflection from
the
left to the left (from the right to the right)
, and $t_{\rm R}$ ($t_{\rm L}$) is an $N \times N$
invertible matrix that describes the transmission
from
the left to the right (from the right to the left)
. In a similar manner, the transfer matrix $M$ is defined by
\begin{equation}
\left( \begin{array}{@{\,}c@{\,}} 
	b_{\rm out}^{+} \\ b_{\rm in}^{-} \\ 
	\end{array} \right) 
= M \left( \begin{array}{@{\,}c@{\,}} 
	a_{\rm in}^{+} \\ a_{\rm out}^{-} \\ 
	\end{array} \right).
\end{equation}
These definitions of the scattering matrix and transfer matrix mean
\begin{equation} \begin{split}
a_{\rm out}^{-} = r_{\rm L} a_{\rm in}^{+} + t_{\rm L} b_{\rm in}^{-}&,\quad
b_{\rm out}^{+} = t_{\rm R} a_{\rm in}^{+} + r_{\rm R} b_{\rm in}^{-}, \\
b_{\rm out}^{+} = M_{11} a_{\rm in}^{+} + M_{12} a_{\rm out}^{-}&,\quad
b_{\rm in}^{-} = M_{21} a_{\rm in}^{+} + M_{22} a_{\rm out}^{-},
\end{split} \end{equation}
which leads to
\begin{equation}
M = \left( \begin{array}{@{\,}cc@{\,}} 
	t_{\rm R} - r_{\rm R} t_{\rm L}^{-1} r_{\rm L} & r_{\rm R} t_{\rm L}^{-1} \\
	- t_{\rm L}^{-1} r_{\rm L} & t_{\rm L}^{-1} \\ 
	\end{array} \right),
\end{equation}
and 
\begin{equation}
r_{\rm L} = - M_{22}^{-1} M_{21},\quad
r_{\rm R} = M_{12} M_{22}^{-1},\quad
t_{\rm L} = M_{22}^{-1},\quad
t_{\rm R} = M_{11} - M_{12} M_{22}^{-1} M_{21}.
\end{equation}
Notably, we have
\begin{equation}
\det M = \det \left[ \left( t_{\rm R} - r_{\rm R} t_{\rm L}^{-1} r_{\rm L} \right) t_{\rm L}^{-1} - \left( r_{\rm R} t_{\rm L}^{-1} \right) t_{\rm L} \left( - t_{\rm L}^{-1} r_{\rm L} \right) t_{\rm L}^{-1} \right]
= \frac{\det t_{\rm R}}{\det t_{\rm L}}.
\end{equation}

When the system is closed (i.e., isolated from the environment) and hence the Hamiltonian $H$ is Hermitian, the amplitude of the waves is conserved under the scattering:
\begin{equation}
\left| a_{\rm in}^{+} \right|^{2} + \left| b_{\rm in}^{-} \right|^{2}
= \left| a_{\rm out}^{-} \right|^{2} + \left| b_{\rm out}^{+} \right|^{2}.
\end{equation}
As a result, the scattering matrix $S$ is unitary:
\begin{equation}
S^{\dag} S = S S^{\dag} = 1.
\end{equation}
Unitarity of $S$ implies that the Hermitian matrices $t_{\rm L} t_{\rm L}^{\dag}$, $t_{\rm R} t_{\rm R}^{\dag}$, $1-r_{\rm L} r_{\rm L}^{\dag}$, and $1-r_{\rm R} r_{\rm R}^{\dag}$ have the same set of eigenvalues. In addition, since we have
\begin{equation}
\left| a_{\rm in}^{+} \right|^{2} - \left| a_{\rm out}^{-} \right|^{2}
= \left| b_{\rm out}^{+} \right|^{2} - \left| b_{\rm in}^{-} \right|^{2}
= \left( \begin{array}{@{\,}c@{\,}} 
	b_{\rm out}^{+} \\ b_{\rm in}^{-} \\
	\end{array} \right)^{\dag} \tau_{3} \left( \begin{array}{@{\,}c@{\,}} 
	b_{\rm out}^{+} \\ b_{\rm in}^{-} \\
	\end{array} \right)
= \left( \begin{array}{@{\,}c@{\,}} 
	a_{\rm in}^{+} \\ a_{\rm out}^{-} \\
	\end{array} \right)^{\dag} M^{\dag} \tau_{3} M \left( \begin{array}{@{\,}c@{\,}} 
	a_{\rm in}^{+} \\ a_{\rm out}^{-} \\
	\end{array} \right)
\end{equation}
with a Pauli matrix $\tau_{3}$, the transfer matrix $M$ is pseudo-unitary:
\begin{equation}
\tau_{3} M^{\dag} \tau_{3}^{-1} = M^{-1}.
\end{equation}
However, when the system exchanges energy or particles with the environment and
hence the Hamiltonian $H$ is non-Hermitian, the scattering matrix $S$ and the
transfer matrix $M$ are not unitary and pseudo-unitary, respectively,
which can change the universality of localization transitions.

\subsection{Symmetry}

\begin{table}[b]
	\centering
	\caption{Symmetry of non-Hermitian Hamiltonians $H$ and nonunitary scattering matrices $S$. A typical representation of symmetry is shown for each class, where $\sigma_{i}$'s and $\tau_{i}$'s are Pauli matrices that describe the spin and valley degrees of freedom, respectively. Furthermore, the type of delocalization and the typical conductances for sufficiently large systems $L \gg \ell$ are shown with non-Hermiticity $\gamma$ and the mean free path $\ell > 0$.}
     \begin{tabular}{ccccc} \hline \hline
     ~~Class~~ & ~~Symmetry of $H$~~ & ~~Symmetry of $S$~~ & ~~~Delocalization~~~ & Typical conductances ($L \gg \ell$)  \\ \hline
     A & No & No & ~~Unidirectional~~ & $e^{( \pm \gamma - 1/\ell ) L}$ \\
     AI & $\tau_{1} H^{*} \tau_{1}^{-1} = H$ & $S^{*} \left( E \right) = S^{-1} \left( E^{*} \right)$ & ~~Unidirectional~~ & $e^{( \pm \gamma - 1/\ell ) L}$ \\
     $\text{AI}^{\dag}$ & $\tau_{1} H^{T} \tau_{1}^{-1} = H$ & $S^{T} \left( E \right) = S \left( E \right)$ & No & $e^{-L/\ell}$ \\
     AII & ~~$\left( \sigma_{2} \tau_{1} \right) H^{*} \left( \sigma_{2} \tau_{1} \right)^{-1} = H$~~ & ~~$\sigma_{2} S^{*} \left( E \right) \sigma_{2}^{-1} = S^{-1} \left( E^{*} \right)$~~ & Unidirectional & $e^{( \pm \gamma - 1/\ell ) L}$ \\
     $\text{AII}^{\dag}$ & ~~$\left( \sigma_{2} \tau_{1} \right) H^{T} \left( \sigma_{2} \tau_{1} \right)^{-1} = H$~~ & ~~$\sigma_{2} S^{T} \left( E \right) \sigma_{2}^{-1} = S \left( E \right)$~~ & Bidirectional & $e^{( \left| \gamma \right| - 1/\ell ) L}$ \\ 
     AIII & $\tau_{1} H^{\dag} \tau_{1}^{-1} = - H$ & $S^{\dag} \left( E \right) = S \left( - E^{*} \right)$ & ~~Bidirectional (chiral unitary)~~ & $e^{-\sqrt{8L/\pi\ell}}$ \\
     $\text{AIII}^{\dag}$ & $\tau_{1} H \tau_{1}^{-1} = - H$ & $S \left( E \right) = S^{-1} \left( - E \right)$ & Unidirectional & $e^{\pm \gamma\ell -\sqrt{8L/\pi\ell}}$ \\ \hline \hline
    \end{tabular}
	\label{Stab: symmetry}
\end{table}

When the system respects symmetry, certain constraints are imposed on the non-Hermitian Hamiltonian $H$ and the nonunitary scattering matrix $S$. Importantly, non-Hermiticity and nonunitarity change the nature of symmetry, and the symmetry constraints become different from the conventional constraints in Hermitian systems. While the symmetry constraints for non-Hermitian Hamiltonians $H$ are identified in Refs.~\cite{BL-02, KSUS-19}, we here provide the symmetry constraints for nonunitary scattering matrices $S$ (Table~\ref{Stab: symmetry}). Our discussions are based on the relationship between $H$ and $S$ (Mahaux-Weidenm\"uller formula~\cite{Beenakker-review, *Beenakker-review-sc}):
\begin{equation}
S \left( E \right) = \frac{1 - \ii \pi K \left( E \right)}{1 + \ii \pi K \left( E \right)},\quad
K \left( E \right) := W^{\dag} \frac{1}{E-H} W,
\end{equation}
where $E \in \mathbb{C}$ is an energy of the incident and the scattered waves, and $W$ describes the coupling between the system and the leads and is assumed to commute with the symmetry operations.

\medskip
\paragraph{Time-reversal symmetry and reciprocity.\,---}Time-reversal symmetry for non-Hermitian Hamiltonians $H$ is defined by 
\begin{equation}
\mathcal{T} H^{*} \mathcal{T}^{-1} = H,
	\label{Seq: TRS}
\end{equation}
where $\mathcal{T}$ is a unitary matrix (i.e., $\mathcal{T}^{\dag} \mathcal{T} = \mathcal{T} \mathcal{T}^{\dag} = 1$) and commutes with the coupling matrix $W$ (i.e., $\mathcal{T} W^{*} \mathcal{T}^{-1} = W$). Then, we have
\begin{equation}
\mathcal{T} K^{*} \left( E \right) \mathcal{T}^{-1}
= W^{\dag} \frac{1}{E^{*} - H} W
= K \left( E^{*} \right),
\end{equation}
and hence
\begin{equation}
\mathcal{T} S^{*} \left( E \right) \mathcal{T}^{-1}
= \frac{1+\ii \pi K \left( E^{*} \right)}{1-\ii \pi K \left( E^{*} \right)}
= S^{-1} \left( E^{*} \right).
	\label{Seq: TRS-S}
\end{equation}
For $\mathcal{T} \mathcal{T}^{*} = +1$ ($\mathcal{T} \mathcal{T}^{*} = -1$), non-Hermitian systems are defined to belong to class AI (AII)~\cite{KSUS-19}.

Non-Hermiticity enables a Hermitian-conjugate counterpart of time-reversal symmetry as another fundamental symmetry ($\text{TRS}^{\dag}$ in Ref.~\cite{KSUS-19}). In fact, because of the difference between complex conjugation and transposition, the symmetry defined by
\begin{equation}
\mathcal{T} H^{T} \mathcal{T}^{-1} = H
	\label{Seq: reciprocity}
\end{equation}
is distinct from time-reversal symmetry defined by Eq.~(\ref{Seq: TRS}). As a consequence of this symmetry, we have
\begin{equation}
\mathcal{T} K^{T} \left( E \right) \mathcal{T}^{-1}
= W^{\dag} \frac{1}{E - H} W
= K \left( E \right),
\end{equation}
and hence
\begin{equation}
\mathcal{T} S^{T} \left( E \right) \mathcal{T}^{-1}
= \frac{1-\ii \pi K \left( E \right)}{1+\ii \pi K \left( E^{*} \right)}
= S \left( E \right).
	\label{Seq: reciprocity-S}
\end{equation}
For $\mathcal{T} \mathcal{T}^{*} = +1$ ($\mathcal{T} \mathcal{T}^{*} = -1$), non-Hermitian systems are defined to belong to class $\text{AI}^{\dag}$ ($\text{AII}^{\dag}$)~\cite{KSUS-19}. For $\mathcal{T} = 1$, for example, Eq.~(\ref{Seq: reciprocity-S}) implies 
\begin{equation}
r_{\rm L}^{T} = r_{\rm L},\quad
r_{\rm R}^{T} = r_{\rm R},\quad
t_{\rm L}^{T} = t_{\rm R},
\end{equation}
and hence the transmission amplitude from
the
right to the left is equivalent to the transmission amplitude
from
the
left to the right [i.e., $\mathrm{tr}\,( t_{\rm L} t_{\rm L}^{\dag} ) = \mathrm{tr}\,( t_{\rm R} t_{\rm R}^{\dag} )$]. Thus, the symmetry defined by Eq.~(\ref{Seq: reciprocity}) physically means reciprocity in non-Hermitian systems. Importantly, time-reversal symmetry and reciprocity are distinct from each other in non-Hermitian systems, although they are equivalent to each other in Hermitian systems; the symmetry constraints in Eqs.~(\ref{Seq: TRS-S}) and (\ref{Seq: reciprocity-S}) and the consequent universality of localization transitions are different from each other.

\medskip
\paragraph{Particle-hole symmetry.\,---}Particle-hole symmetry for non-Hermitian Hamiltonians $H$ is defined by
\begin{equation}
\mathcal{C} H^{T} \mathcal{C}^{-1} = - H,
\end{equation}
where $\mathcal{C}$ is a unitary matrix and commutes with the coupling matrix $W$ (i.e., $\mathcal{C} W^{T} \mathcal{C}^{-1} = W$). Then, we have
\begin{equation}
\mathcal{C} K^{T} \left( E \right) \mathcal{C}^{-1}
= W^{\dag} \frac{1}{E + H} W
= - K \left( -E \right),
\end{equation}
and hence
\begin{equation}
\mathcal{C} S^{T} \left( E \right) \mathcal{C}^{-1}
= \frac{1+\ii \pi K \left( -E \right)}{1-\ii \pi K \left( -E \right)}
= S^{-1} \left( -E \right).
	\label{Seq: PHS-S}
\end{equation}
In a similar manner to time-reversal symmetry and reciprocity, non-Hermiticity enables a Hermitian-conjugate counterpart of particle-hole symmetry ($\text{PHS}^{\dag}$ in Ref.~\cite{KSUS-19}):
\begin{equation}
\mathcal{C} H^{*} \mathcal{C}^{-1} = - H.
\end{equation}
Then, we have
\begin{equation}
\mathcal{C} K^{*} \left( E \right) \mathcal{C}^{-1}
= W^{\dag} \frac{1}{E^{*} + H} W
= - K \left( -E^{*} \right),
\end{equation}
and hence
\begin{equation}
\mathcal{C} S^{*} \left( E \right) \mathcal{C}^{-1}
= \frac{1-\ii \pi K \left( -E^{*} \right)}{1+\ii \pi K \left( -E^{*} \right)}
= S \left( -E^{*} \right).
	\label{Seq: PHS+-S}
\end{equation}
Whereas Eqs.~(\ref{Seq: PHS-S}) and (\ref{Seq: PHS+-S}) are equivalent to each other in the presence of unitarity, they are not in the nonunitary case.

\medskip
\paragraph{Chiral symmetry and sublattice symmetry.\,---}Chiral symmetry for non-Hermitian Hamiltonians $H$ is defined by
\begin{equation}
\Gamma H^{\dag} \Gamma^{-1} = -H,
\end{equation}
where $\Gamma$ is a unitary and Hermitian matrix and commutes with the coupling matrix $W$ (i.e., $\Gamma W \Gamma^{-1} = W$). Then, we have
\begin{equation}
\Gamma K^{\dag} \left( E \right) \Gamma^{-1}
= W^{\dag} \frac{1}{E^{*} + H} W
= - K \left( -E^{*} \right),
\end{equation}
and hence
\begin{equation}
\Gamma S^{\dag} \left( E \right) \Gamma^{-1}
= \frac{1+\ii \pi K \left( -E^{*} \right)}{1-\ii \pi K \left( -E^{*} \right)}
= S \left( -E^{*} \right).
	\label{Seq: CS-S}
\end{equation}
On the other hand, sublattice symmetry is defined by
\begin{equation}
\mathcal{S} H \mathcal{S}^{-1} = -H,
\end{equation}
where $\mathcal{S}$ is a unitary and Hermitian matrix and commutes with the
coupling matrix $W$.
Then, we have
\begin{equation}
\mathcal{S} K \left( E \right) \mathcal{S}^{-1}
= W^{\dag} \frac{1}{E + H} W
= - K \left( -E \right),
\end{equation}
and hence
\begin{equation}
\mathcal{S} S \left( E \right) \mathcal{S}^{-1}
= \frac{1+\ii \pi K \left( -E \right)}{1-\ii \pi K \left( -E \right)}
= S^{-1} \left( -E \right).
	\label{Seq: SLS-S}
\end{equation}
Whereas chiral symmetry defined by Eq.~(\ref{Seq: CS-S}) and sublattice symmetry defined by Eq.~(\ref{Seq: SLS-S}) are equivalent to each other in the presence of unitarity, they are not in the nonunitary case. Non-Hermitian systems that respect chiral symmetry in Eq.~(\ref{Seq: CS-S}) [sublattice symmetry in Eq.~(\ref{Seq: SLS-S})] are defined to belong to class AIII ($\text{AIII}^{\dag}$)~\cite{KSUS-19}.

\section{SII.~Scattering theory in one dimension}
	\label{Ssec: Born approximation}

We provide a scattering theory of non-Hermitian Hamiltonians in one dimension. For the non-Hermitian Hamiltonian $H \left( x \right) = H_{0} \left( x \right) + V \left( x \right)$, 
let $E$ and $\varphi \left( x \right)$ be
an eigenenergy and the corresponding right eigenstate, respectively:
\begin{equation}
\left[ H_{0} \left( x \right) + V \left( x \right) \right] \varphi \left( x \right)
= E\,\varphi \left( x \right).
\end{equation}
For this Schr\"odinger equation, we define the Green's function $G_{0} \left( x \right)$ by
\begin{equation}
H_{0} \left( x \right) G_{0} \left( x \right) + \delta \left( x \right)
= E\,G_{0} \left( x \right).
	\label{Seq: Green-def}
\end{equation}
Then, the eigenstate $\varphi \left( x\right)$ satisfies
\begin{equation}
\varphi \left( x\right)
= \varphi_{0} \left( x\right) + \int_{-\infty}^{\infty} dy~G_{0} \left( x-y \right) V \left( y \right) \varphi \left( y \right),
\end{equation}
where $\varphi_{0} \left( x \right)$ is a solution to the Schr\"odinger equation in the absence of the potential $V \left( x \right)$ [i.e., $H_{0} \left( x \right) \varphi_{0} \left( x \right) = E\,\varphi_{0} \left( x \right)$]. For a sufficiently weak potential $V \left( x \right)$, the Born approximation is justified. Up to the second-order Born approximation, the eigenstate is given as $\varphi \left( x \right) \simeq \varphi_{0} \left( x \right) + \varphi_{1} \left( x \right) + \varphi_{2} \left( x \right)$ with
\begin{eqnarray}
\varphi_{1} \left( x \right) &:=& \int_{-\infty}^{\infty} dy~G_{0} \left( x-y \right) V \left( y \right) \varphi_{0} \left( y \right), \label{Seq: Born-1} \\
\varphi_{2} \left( x \right) &:=& \int_{-\infty}^{\infty} dy \int_{-\infty}^{\infty} dz~G_{0} \left( x-y \right) V \left( y \right) G_{0} \left( y-z \right) V \left( z \right) \varphi_{0} \left( z \right). \label{Seq: Born-2}
\end{eqnarray}

In particular, we consider the scattering for $H_{0} \left( x\right) = -\ii
\tau_{3} \partial_{x}$. Here, $\tau_{3}$ is a Pauli matrix that describes
the two valley degrees of freedom.
Performing the Fourier transformations
\begin{equation}
G_{0} \left( x \right) = \int_{-\infty}^{\infty} \frac{dk}{2\pi} \tilde{G}_{0} \left( k \right) e^{\ii kx},\quad
\delta \left( x \right) = \int_{-\infty}^{\infty} \frac{dk}{2\pi} e^{\ii kx}
\end{equation}
for Eq.~(\ref{Seq: Green-def}), we have
\begin{equation}
\tilde{G}_{0} \left( k \right) 
= \left( E - k\tau_{3} \right)^{-1}
= \left( \begin{array}{@{\,}cc@{\,}} 
	1/ \left( E-k\right) & 0 \\
	0 & 1/ \left( E+k\right) \\ 
	\end{array} \right).
\end{equation}
Using the formulas
\begin{equation}
\int_{-\infty}^{\infty} \frac{dk}{2\pi} \frac{e^{\ii kx}}{E \pm \ii \varepsilon - k}
= \mp \ii e^{\ii Ex} \theta \left( \pm x \right),\quad
\int_{-\infty}^{\infty} \frac{dk}{2\pi} \frac{e^{\ii kx}}{E \pm \ii \varepsilon + k}
= \mp \ii e^{-\ii Ex} \theta \left( \mp x \right),
\end{equation}
with a positive infinitesimal constant $\varepsilon > 0$ and the Heaviside step function $\theta$, we have
\begin{equation}
G_{0} \left( x; E \pm \ii \varepsilon \right)
= \left( \begin{array}{@{\,}cc@{\,}} 
	\mp \ii e^{\ii Ex} \theta \left( \pm x \right) & 0 \\
	0 & \mp \ii e^{-\ii Ex} \theta \left( \mp x \right) \\ 
	\end{array} \right).
\end{equation}

\section{SIII.~Scaling equations (functional renormalization group equations)}

We formulate the scaling equations (functional renormalization group equations) for non-Hermitian disordered systems in one dimension. Our formulation is based on the random-matrix approach developed for Hermitian quasi-one-dimensional systems by Dorokhov, and by Mello, Pereyra, and Kumar~\cite{Dorokhov-82, Mello-88, Beenakker-review, *Beenakker-review-sc}. The conductance from the left to the right (from the right to the left) is given by the sum of the transmission eigenvalues from the left to the right (from the right to the left) according to the Landauer formula~\cite{Datta-textbook}. Then, we consider the incremental change of the transmission eigenvalues upon attachment of a thin slice of length $dL$ to the system of length $L$. The transmission matrix and the reflection matrix are respectively defined as $t_{\rm L/R} \left( L\right)$ and $r_{\rm L/R} \left( L\right)$ for the original system and $t_{\rm L/R} \left( dL\right)$ and $r_{\rm L/R} \left( dL\right)$ for the attached
  thin slice. The transmission matrix and the reflection matrix of the combined system of length $L+dL$ are given as~\cite{Datta-textbook}
\begin{eqnarray}
t_{\rm R} \left( L+dL \right)
&=& t_{\rm R} \left( L \right) \left[ 1 - r_{\rm R} \left( dL \right) r_{\rm L} \left( L \right) \right]^{-1} t_{\rm R} \left( dL \right), \label{Seq: combination-t} \\
r_{\rm L} \left( L+dL \right)
&=& r_{\rm L} \left( dL \right) + t_{\rm L} \left( dL \right) \left[ 1 - r_{\rm L} \left( L \right) r_{\rm R} \left( dL \right) \right]^{-1} r_{\rm L} \left( L \right) t_{\rm R} \left( dL \right). \label{Seq: combination-r}
\end{eqnarray}
The scattering in the thin slice can be treated perturbatively (i.e., by the Born approximation summarized in Sec.~SII) for sufficiently weak disorder such that the mean free path $\ell$ is much smaller than the Fermi wavelength. Moreover, the incident wave is assumed to be independently and uniformly distributed in the parameter space determined by symmetry. After the above calculations, we have the moments of the transmission eigenvalues and the reflection eigenvalues, which result in the Fokker-Planck equation (DMPK equation) of their probability distribution. This probability distribution provides all the information about the transmission eigenvalues and the conductances. In Hermitian systems, we have $t_{\rm R} t_{\rm R}^{\dag} + r_{\rm L} r_{\rm L}^{\dag} = 1$ as a direct result of unitarity of scattering matrices, and the Fokker-Planck equation can be described solely by the transmission eigenvalues. In non-Hermitian systems, by contrast, the transmission eigenvalues are independent of the reflection eigenvalues, and hence the Fokker-Planck equation is described by both of them. The types of the delocalization and the typical conductances for $L \gg \ell$ are summarized in Table~\ref{Stab: symmetry}.

\subsection{Class A}

We investigate the following non-Hermitian continuum model with disorder:
\begin{equation}
h = \left( -\ii \partial_{x} + \ii \gamma_{3}/2\right) \tau_{3} + m_{0} \left( x \right) + \left( m_{1} \left( x \right) + \ii \gamma_{1}/2 \right) \tau_{1} + \left( m_{2} \left( x \right) + \ii \gamma_{2}/2 \right) \tau_{2},
	\label{Seq: Hamiltonian - A}
\end{equation}
where $\gamma_{i}$'s ($i=1,2,3$) are the degrees of non-Hermiticity, and
$\tau_{i}$'s are Pauli matrices
that describe
the valley degrees of freedom.
The disorder is defined to satisfy
\begin{equation}
\braket{m_{i} \left( x \right)} = 0,\quad
\braket{m_{i} \left( x \right) m_{j} \left( x' \right)} = 2\mu_{i} \delta_{ij} \delta \left( x-x' \right),
\end{equation}
where the bracket $\braket{\star}$ denotes the ensemble average. This continuous model describes a generic non-Hermitian wire having a single channel with two valleys. For lattice models, $m_{0} \left( x \right)$ and $m_{1} \left( x \right)$ correspond to disordered onsite potential, while $m_{2} \left( x \right)$ corresponds to disordered hopping. A constant imaginary term such as $\ii \gamma_{0}/2$ is omitted since it does not affect the localization of eigenstates. The Hatano-Nelson model~\cite{Hatano-Nelson-96, *Hatano-Nelson-97, *Hatano-Nelson-98} at the band center (i.e., $\mathrm{Re}\,E = 0$) is described by $m_{2} = \gamma_{1} = \gamma_{2} = 0$, and the asymmetry of the hopping amplitudes corresponds to non-Hermiticity $\gamma_{3}$.

We begin with solving a scattering problem for a thin slice of length $dL$. Suppose that an incident wave $e^{\ii kx} \ket{+}$ enters the thin slice at $\left[ 0, dL \right]$ from the left, where $\ket{\pm}$ is the eigenstate of $\tau_{3}$ with the eigenvalue $\pm 1$. The incident wave satisfies $-\ii \tau_{3} \partial_{x} \left( e^{\ii kx} \ket{+} \right) = k\ket{+}$ and is indeed a right-moving wave with the eigenenergy $E=k$. In the following, we assume $E = k \in \mathbb{R}$ so that the incident wave will be a plane wave. For $x<0$, up to the second-order Born approximation in Eqs.~(\ref{Seq: Born-1}) and (\ref{Seq: Born-2}), the eigenstate is given as $\varphi \left( x \right) \simeq e^{\ii kx} \ket{+} + \varphi_{1} \left( x \right) + \varphi_{2} \left( x \right)$ with
\begin{eqnarray}
\varphi_{1} \left( x \right) 
&=& \int_{0}^{dL} dy \left( \begin{array}{@{\,}cc@{\,}} 
	0 & 0 \\ 0 & -\ii e^{-\ii k \left( x-y \right)} \\
	\end{array} \right) \left( \begin{array}{@{\,}cc@{\,}} 
	m_{0} \left( y \right) + \ii \gamma_{3}/2 & m_{1} \left( y \right) - \ii m_{2} \left( y \right) + \ii \gamma_{1}/2  + \gamma_{2}/2 \\ m_{1} \left( y \right) + \ii m_{2} \left( y \right) + \ii \gamma_{1}/2 - \gamma_{2}/2 & m_{0} \left( y \right) - \ii \gamma_{3}/2
	\end{array} \right) e^{\ii ky} \ket{+} \nonumber \\
&=& -\ii e^{-\ii kx} \int_{0}^{dL} dy~e^{2\ii ky} \left( m_{1} \left( y \right) + \ii m_{2} \left( y \right) + \ii \gamma_{1}/2 - \gamma_{2}/2\right) \ket{-} \nonumber \\
&\simeq& -\ii \left[ m_{1} \left( dL/2 \right) + \ii m_{2} \left( dL/2 \right) + \ii \gamma_{1}/2 - \gamma_{2}/2 \right] \left( dL \right) e^{-\ii kx} \ket{-},
\end{eqnarray}
\begin{eqnarray}
\varphi_{2} \left( x \right)
&=& -e^{-\ii kx} \int_{0}^{dL} dy \int_{0}^{dL} dz~e^{\ii k \left( y+z \right)} \left( \begin{array}{@{\,}cc@{\,}} 
	0 & 0 \\ m_{1} \left( y \right) + \ii m_{2} \left( y \right) + \ii \gamma_{1}/2 - \gamma_{2}/2 & m_{0} \left( y \right) - \ii \gamma_{3}/2 \\
	\end{array} \right) \nonumber \\
&&\qquad\qquad\qquad\times \left( \begin{array}{@{\,}cc@{\,}} 
	e^{\ii k \left( y-z \right)} \theta \left( y-z \right) & 0 \\ 0 & e^{\ii k \left( z-y \right)} \theta \left( z-y \right) \\
	\end{array} \right)
	\left( \begin{array}{@{\,}c@{\,}} 
	m_{0} \left( z \right) + \ii \gamma_{3}/2 \\ m_{1} \left( z \right) + \ii m_{2} \left( y \right) + \ii \gamma_{1}/2 - \gamma_{2}/2 \\
	\end{array} \right) 
\simeq 0.
\end{eqnarray}
On the other hand, for $x > dL$, we have
\begin{eqnarray}
\varphi_{1} \left( x \right) 
&=& \int_{0}^{dL} dy \left( \begin{array}{@{\,}cc@{\,}} 
	-\ii e^{\ii k \left( x-y \right)} & 0 \\ 0 & 0 \\
	\end{array} \right) \left( \begin{array}{@{\,}cc@{\,}} 
	m_{0} \left( y \right) + \ii \gamma_{3}/2 & m_{1} \left( y \right) - \ii m_{2} \left( y \right) + \ii \gamma_{1}/2  + \gamma_{2}/2 \\ m_{1} \left( y \right) + \ii m_{2} \left( y \right) + \ii \gamma_{1}/2 - \gamma_{2}/2 & m_{0} \left( y \right) - \ii \gamma_{3}/2
	\end{array} \right) e^{\ii ky} \ket{+} \nonumber \\
&=& -\ii e^{\ii kx} \int_{0}^{dL} dy \left( m_{0} \left( y \right) + \ii \gamma_{3}/2 \right) \ket{+} \nonumber \\
&\simeq& -\ii \left[ m_{0} \left( dL/2 \right) + \ii \gamma_{3}/2 \right] \left( dL \right) e^{\ii kx} \ket{+},
\end{eqnarray} 
\begin{eqnarray}
\varphi_{2} \left( x \right)
&=& -e^{\ii kx} \int_{0}^{dL} dy \int_{0}^{dL} dz~e^{\ii k \left( -y+z \right)} \left( \begin{array}{@{\,}cc@{\,}} 
	m_{0} \left( y \right) + \ii \gamma_{3}/2 & m_{1} \left( y \right) - \ii m_{2} \left( y \right) + \ii \gamma_{1}/2 + \gamma_{2}/2 \\ 0 & 0 \\
	\end{array} \right) \nonumber \\
&&\qquad\qquad\times \left( \begin{array}{@{\,}cc@{\,}} 
	e^{\ii k \left( y-z \right)} \theta \left( y-z \right) & 0 \\ 0 & e^{\ii k \left( z-y \right)} \theta \left( z-y \right) \\
	\end{array} \right)
	\left( \begin{array}{@{\,}c@{\,}} 
	m_{0} \left( z \right) + \ii \gamma_{3}/2 \\ m_{1} \left( z \right) + \ii m_{2} \left( y \right) + \ii \gamma_{1}/2 - \gamma_{2}/2 \\
	\end{array} \right) \nonumber \\
&\simeq& -e^{\ii kx} \int_{0}^{dL} dy \int_{0}^{dL} dz \left[ e^{\ii k \left( y-z \right)} \theta \left( y-z \right) m_{0} \left( y \right) m_{0} \left( z \right) + e^{\ii k \left( z-y \right)} \theta \left( z-y \right) \left( m_{1} \left( y \right) m_{1} \left( z \right) + m_{2} \left( y \right) m_{2} \left( z \right) \right) \right] \ket{+} \nonumber \\
&\simeq& - \frac{1}{2} \left[ \left( m_{0} \left( dL/2 \right) \right)^{2} + \left( m_{1} \left( dL/2 \right) \right)^{2} + \left( m_{2} \left( dL/2 \right) \right)^{2} \right] \left( dL \right)^{2} e^{\ii kx} \ket{+}.
\end{eqnarray}
The above results mean 
\begin{eqnarray}
r_{\rm L} \left( dL \right) &\simeq& -\ii \left( m_{1} + \ii m_{2} + \ii \gamma_{1}/2 - \gamma_{2}/2 \right) dL,\\
t_{\rm R} \left( dL \right) &\simeq& 1 - \ii \left( m_{0} - E + \ii \gamma_{3}/2 \right) \left( dL \right) - \frac{1}{2} \left( m_{0}^{2} + m_{1}^{2} + m_{2}^{2} \right) \left( dL \right)^{2}.
\end{eqnarray}
Similarly, for a left-moving incident wave $e^{-\ii kx} \ket{-}$ that enters the system from the right, $r_{\rm R} \left( dL \right)$ and $t_{\rm L} \left( dL \right)$ are given as
\begin{eqnarray}
r_{\rm R} \left( dL \right) &\simeq& -\ii \left( m_{1} - \ii m_{2} + \ii \gamma_{1}/2 + \gamma_{2}/2 \right) dL,\\
t_{\rm L} \left( dL \right) &\simeq& 1 - \ii \left( m_{0} - E - \ii \gamma_{3}/2 \right) \left( dL \right) - \frac{1}{2} \left( m_{0}^{2} + m_{1}^{2} + m_{2}^{2} \right) \left( dL \right)^{2}.
\end{eqnarray}
Thus, we have
\begin{equation} \begin{split}
\braket{\left| r_{\rm L} \left( dL \right) \right|^{2}}
= 2 \left( \mu_{1} + \mu_{2} \right) dL,\quad
\braket{\left| t_{\rm R} \left( dL \right) \right|^{2}}
= 1 - 2 \left( \mu_{1} + \mu_{2} - \gamma_{3}/2 \right) dL; \\
\braket{\left| r_{\rm R} \left( dL \right) \right|^{2}}
= 2 \left( \mu_{1} + \mu_{2} \right) dL,\quad
\braket{\left| t_{\rm L} \left( dL \right) \right|^{2}}
= 1 - 2 \left( \mu_{1} + \mu_{2} + \gamma_{3}/2 \right) dL.
	\label{Seq: dL - average}
\end{split} \end{equation}
In the absence of non-Hermiticity (i.e., $\gamma_{i} = 0$), we indeed have $\braket{\left| r_{\rm L} \left( dL \right) \right|^{2}} + \braket{\left| t_{\rm R} \left( dL \right) \right|^{2}} = \braket{\left| r_{\rm R} \left( dL \right) \right|^{2}} + \braket{\left| t_{\rm L} \left( dL \right) \right|^{2}} = 1$, which means conservation of currents; however, it is broken by non-Hermiticity $\gamma_{3}$. In the following, we define the mean free path $\ell$ by
\begin{equation}
\braket{\left| r_{\rm L} \left( dL \right) \right|^{2}} = \braket{\left| r_{\rm R} \left( dL \right) \right|^{2}} =: \frac{dL}{\ell},
\quad\mathrm{i.e.},\quad
\ell := \frac{1}{2 \left( \mu_{1} + \mu_{2} \right)}.
\end{equation}

Now, we consider combining the system of length $L$ and the thin slice of length $dL$. Using Eq.~(\ref{Seq: combination-t}), as well as
\begin{equation}
\left| t_{\rm R} \left( dL \right) \right|^{2}
\simeq 1 + \gamma_{3}\,dL - \left( m_{1}^{2} + m_{2}^{2} \right) \left( dL\right)^{2} 
\end{equation}
and 
\begin{eqnarray}
\left| 1 - r_{\rm R} \left( dL \right) r_{\rm L} \left( L \right) \right|^{-2}
&\simeq& 1 + 2 \sqrt{R_{\rm L}} \left[ \left( m_{1} + \gamma_{2}/2\right) \sin \varphi_{\rm L} - \left( m_{2} - \gamma_{1}/2 \right) \cos \varphi_{\rm L} \right] dL \nonumber \\
&&\qquad\qquad\qquad\qquad + \left[ m_{1}^{2} \left( 4 \sin^{2} \varphi_{\rm L} - 1 \right) + m_{2}^{2} \left( 4 \cos^{2} \varphi_{\rm L} - 1 \right) \right] R_{\rm L} \left( dL \right)^{2}
\end{eqnarray}
with $r_{\rm L} =: \sqrt{R_{\rm L}} e^{\ii \varphi_{\rm L}}$, we have
\begin{eqnarray}
\frac{dT_{\rm R}}{T_{\rm R}} 
&\simeq& \left\{ 2 \sqrt{R_{\rm L}} \left[ \left( m_{1} + \gamma_{2}/2\right) \sin \varphi_{\rm L} - \left( m_{2} - \gamma_{1}/2 \right) \cos \varphi_{\rm L} \right]  + \gamma_{3} \right\} dL \nonumber \\
&&\qquad\qquad\qquad\qquad + \left\{ m_{1}^{2} \left[ \left( 4\sin^{2} \varphi_{\rm L} - 1\right) R_{\rm L}- 1 \right] + m_{2}^{2} \left[ \left( 4\cos^{2} \varphi_{\rm L} - 1\right) R_{\rm L}- 1 \right]\right\} \left( dL \right)^{2}.
\end{eqnarray}
This leads to 
\begin{equation}
\frac{\braket{dT_{\rm R}}}{dL}
= \gamma_{3} T_{\rm R} - \frac{T_{\rm R} \left( 1-R_{\rm L}\right)}{\ell},\quad
\frac{\braket{\left( dT_{\rm R} \right)^{2}}}{dL}
= \frac{2T_{\rm R}^{2} R_{\rm L}}{\ell},
	\label{Seq: scaling-TR}
\end{equation}
and the higher moments vanish to the first order in $dL$. Here, the ensemble average $\braket{\star}$ is taken for given $T_{\rm R}$ and $R_{\rm L}$ in two steps, first averaging over the attached thin slice $m_{i}$ and then over the phase $\varphi_{\rm L}$ of the reflected wave. Since $\varphi_{\rm L}$ is assumed to be independently and uniformly distributed over $\left[ 0, 2\pi \right]$, the equations $\braket{\cos \varphi_{\rm L}} = \braket{\sin \varphi_{\rm L}} = 0$ and $\braket{\cos^{2} \varphi_{\rm L}} = \braket{\sin^{2} \varphi_{\rm L}} = 1/2$ are used. We have similar scaling equations also for $T_{\rm L}$ by reversing the sign of the non-Hermiticity $\gamma_{3}$. Moreover, using Eq.~(\ref{Seq: combination-r}), we have
\begin{equation}
\frac{\braket{dR_{\rm L}}}{dL}
= \frac{\left( 1 - R_{\rm L} \right)^{2}}{\ell},\quad
\frac{\braket{\left( dR_{\rm L} \right)^{2}}}{dL}
= \frac{2 R_{\rm L} \left( 1-R_{\rm L} \right)^{2}}{\ell},
	\label{Seq: scaling-RL}
\end{equation}
and the same scaling equations for $R_{\rm R}$.

In the obtained scaling equations~(\ref{Seq: scaling-TR}) and (\ref{Seq: scaling-RL}), non-Hermiticity appears solely through the $\gamma_{3}$ terms. By contrast, the $\gamma_{1}$ and $\gamma_{2}$ terms just shift the phase of the waves and have no influence on the conductances. Consequently, when we define $\tilde{T}_{\rm R}$ and $\tilde{T}_{\rm L}$ by
\begin{equation}
\tilde{T}_{\rm R} := e^{-\gamma_{3} L} T_{\rm R},\quad
\tilde{T}_{\rm L} := e^{+\gamma_{3} L} T_{\rm L},
\end{equation}
the transfer amplitudes $\tilde{T}_{\rm R}$ and $\tilde{T}_{\rm L}$ and the reflection amplitudes $R_{\rm L}$ and $R_{\rm R}$ are described by the conventional Fokker-Planck equation (DMPK equation) for Hermitian systems. In particular, the average conductance $\tilde{G}^{\rm av}$ and the typical conductance $\tilde{G}^{\rm typ}$ are given as~\cite{Dorokhov-82, Mello-88, Beenakker-review, *Beenakker-review-sc}
\begin{equation} \begin{split}
\tilde{G}^{\rm av}/G_{\rm c} &:= \braket{\tilde{G}}/G_{\rm c} = e^{-L/4\ell} f \left( L/\ell \right) \sim e^{-L/4\ell}\quad\left( L/\ell \to \infty \right), \\
\tilde{G}^{\rm typ}/G_{\rm c} &:= e^{\braket{\log \tilde{G}/G_{\rm c}}}
= e^{-L/\ell},
\end{split} \end{equation}
where $G_{\rm c}$ is the conductance quantum, and $f$ is the following slowly-varying function:
\begin{equation}
f \left( x \right) := \frac{2}{\pi} \int_{0}^{\infty} t\,\frac{\tanh t}{\cosh t}\,e^{-xt^2/\pi^2} dt.
\end{equation} 
Thus, the conductances of the original non-Hermitian system are given as
\begin{equation} \begin{split}
G^{\rm av}_{\rm R}/G_{\rm c} \sim e^{\left( \gamma_{3} - 1/4\ell\right) L},\quad
G^{\rm typ}_{\rm R}/G_{\rm c} = e^{\left( \gamma_{3} - 1/\ell\right) L}; \\
G^{\rm av}_{\rm L}/G_{\rm c} \sim e^{\left( -\gamma_{3} - 1/4\ell\right) L},\quad
G^{\rm typ}_{\rm L}/G_{\rm c} = e^{\left( -\gamma_{3} - 1/\ell\right) L}.
\end{split} \end{equation}
Either $G_{\rm R}$ or $G_{\rm L}$ diverges for sufficiently strong non-Hermiticity as a signature of the unidirectional delocalization. For $\gamma_{3} \geq 0$, for example, the transition point at which the typical conductance $G_{\rm R}^{\rm typ}$ from the left to the right begins to diverge is given by $\gamma_{3} = \gamma_{\rm c} := 1/\ell$. Around this transition point, the conductance exhibits the critical behavior $| G_{\rm R}^{\rm typ} - G_{\rm c} |/G_{\rm c} \propto \left| \gamma_{3} - \gamma_{\rm c} \right|$.

\subsection{Classes AI and $\text{AI}^{\dag}$}

Symmetry imposes constraints on systems and can change the universality of localization transitions. Non-Hermitian Hamiltonians in class AI respect time-reversal symmetry. In the presence of time-reversal symmetry defined by $\tau_{1} h^{*} \tau_{1}^{-1} = h$, the non-Hermitian terms $\ii \left( \gamma_{1}/2 \right) \tau_{1}$ and $\ii \left( \gamma_{2}/2 \right) \tau_{2}$ in Eq.~(\ref{Seq: Hamiltonian - A}) disappear. Still, the non-Hermitian term $\ii \left( \gamma_{3}/2 \right) \tau_{3}$ is allowed to be present, which leads to the unidirectional delocalization. Thus, the universality of the localization transitions in class AI is the same as that in class A.

On the other hand, when non-Hermitian Hamiltonians belong to class $\text{AI}^{\dag}$ (orthogonal class) and respect reciprocity defined by $\tau_{1} h^{T} \tau_{1}^{-1} = h$, the non-Hermitian term $\ii \left( \gamma_{3}/2 \right) \tau_{3}$ in Eq.~(\ref{Seq: Hamiltonian - A}) disappears. Consequently, the unidirectional delocalization is forbidden and the conductances for $L \gg \ell$ are given as
\begin{equation}
G^{\rm av}_{\rm R}/G_{\rm c} = G^{\rm av}_{\rm L}/G_{\rm c} \sim e^{- L/4\ell},\quad
G^{\rm typ}_{\rm R}/G_{\rm c} = G^{\rm typ}_{\rm L}/G_{\rm c} = e^{- L/\ell}.
\end{equation}
The universality of the non-Hermitian localization in class $\text{AI}^{\dag}$ is the same as the Hermitian counterpart.

\subsection{Classes AII and $\text{AII}^{\dag}$}

An important feature of class AII (symplectic class) in Hermitian systems is Kramers degeneracy of eigenenergies. This Kramers-pair structure survives even in non-Hermitian systems: eigenstates with real eigenenergies form Kramers pairs in class AII~\cite{Kawabata-19}, whereas generic eigenstates with complex eigenenergies form Kramers pairs in class $\text{AII}^{\dag}$~\cite{KSUS-19}. This difference in the Kramers-pair structure makes a difference in the universality of localization transitions, as described below. It is also notable that the transmission eigenvalues are not generally degenerate in the presence of non-Hermiticity, whereas they form Kramers pairs in Hermitian systems in class AII.

We investigate a non-Hermitian continuum model 
\begin{equation}
h = \left( \ii \partial_{x} + \Delta \sigma_{1} + \ii \gamma_{03}/2 \right) \tau_{3} + m_{0} \left( x \right) + m_{1} \left( x \right) \tau_{1},
\end{equation} 
which respects time-reversal symmetry $\left( \sigma_{2}\tau_{1} \right) h^{*}
\left( \sigma_{2}\tau_{1} \right)^{-1} = h$ and hence belongs to class AII.
Here, Pauli matrices $\sigma_{i}$'s describe the internal degrees of freedom
such as spin,
while $\tau_{i}$'s describe
the valley degrees of freedom.
The non-Hermitian terms such as $\ii \left( \gamma_{13}/2\right) \sigma_{1} \tau_{3}$, $\ii \left( \gamma_{23}/2\right) \sigma_{2} \tau_{3}$, and $\ii \left( \gamma_{33}/2\right) \sigma_{3} \tau_{3}$ are forbidden because of time-reversal symmetry. In a similar manner to class A, the reflection and the transmission matrices of a thin slice of the system are given as
\begin{eqnarray}
r_{\rm L} \left( dL \right)
&=& r_{\rm R} \left( dL \right)
= -\ii m_{1} dL,\\
t_{\rm R} \left( dL \right)
&=& 1 - \ii \left( m_{0} - E + \Delta \sigma_{1} + \ii \gamma_{03}/2 \right) dL - \frac{1}{2} \left( m_{0}^{2} + m_{1}^{2} \right) \left( dL \right)^{2}, \\
t_{\rm L} \left( dL \right)
&=& 1 - \ii \left( m_{0} - E - \Delta \sigma_{1} - \ii \gamma_{03}/2 \right) dL - \frac{1}{2} \left( m_{0}^{2} + m_{1}^{2} \right) \left( dL \right)^{2}.
\end{eqnarray}
Thus, we have
\begin{equation} \begin{split}
\frac{1}{2} \braket{ \mathrm{tr}\,[ r_{\rm L} \left( dL \right) r_{\rm L}^{\dag} \left( dL \right) ]}
&= \frac{1}{2} \braket{ \mathrm{tr}\,[ r_{\rm R} \left( dL \right) r_{\rm R}^{\dag} \left( dL \right) ]}
= \frac{dL}{\ell},\\
\frac{1}{2} \braket{ \mathrm{tr}\,[ t_{\rm R} \left( dL \right) t_{\rm R}^{\dag} \left( dL \right) ]}
= 1 - \left( \frac{1}{\ell} - \gamma_{03} \right) dL&,\quad
\frac{1}{2} \braket{ \mathrm{tr}\,[ t_{\rm L} \left( dL \right) t_{\rm L}^{\dag} \left( dL \right) ]}
= 1 - \left( \frac{1}{\ell} + \gamma_{03} \right) dL,
\end{split} \end{equation}
with the mean free path $\ell := 1/2\mu_{1}$. Similarly to Eq.~(\ref{Seq: dL - average}) for class A, one of $\braket{ \mathrm{tr}\,[ t_{\rm R} \left( dL \right) t_{\rm R}^{\dag} \left( dL \right) ]}$ and $\braket{ \mathrm{tr}\,[ t_{\rm L} \left( dL \right) t_{\rm L}^{\dag} \left( dL \right) ]}$ is amplified by non-Hermiticity $\gamma_{03}$ and the other is attenuated. Hence, the same scaling equations [i.e., Eqs.~(\ref{Seq: scaling-TR}) and (\ref{Seq: scaling-RL})] describe the probability distribution of the conductances, and either of the conductances $G_{\rm R}$ and $G_{\rm L}$ is amplified by non-Hermiticity $\gamma_{03}$. Thus, the unidirectional delocalization is realized in the same manner as class A.

By contrast, a different type of non-Hermitian delocalization appears in class $\text{AII}^{\dag}$. We investigate a non-Hermitian continuum model 
\begin{equation}
h = \left( \ii \partial_{x} + \Delta \sigma_{1} + \ii \left( \gamma_{33}/2 \right) \sigma_{3} \right) \tau_{3} + m_{0} \left( x \right) + m_{1} \left( x \right) \tau_{1},
\end{equation} 
which respects reciprocity $\left( \sigma_{2}\tau_{1} \right) h^{T} \left( \sigma_{2}\tau_{1} \right)^{-1} = H$ and hence belongs to class $\text{AII}^{\dag}$. In contrast to class AII, the non-Hermitian term $\ii \left( \gamma_{33}/2 \right) \sigma_{3} \tau_{3}$ is allowed to be present, whereas $\ii \left( \gamma_{03}/2 \right) \tau_{3}$ is forbidden. In this case, the reflection and the transmission matrices of a thin slice are given as
\begin{eqnarray}
r_{\rm L} \left( dL \right)
&=& r_{\rm R} \left( dL \right)
= -\ii m_{1} dL,\\
t_{\rm R} \left( dL \right)
&=& 1 - \ii \left( m_{0} - E + \Delta \sigma_{1} + \ii \left( \gamma_{33}/2 \right) \sigma_{3} \right) dL - \frac{1}{2} \left( m_{0}^{2} + m_{1}^{2} \right) \left( dL \right)^{2}, \\
t_{\rm L} \left( dL \right)
&=& 1 - \ii \left( m_{0} - E - \Delta \sigma_{1} - \ii \left( \gamma_{33}/2 \right) \sigma_{3} \right) dL - \frac{1}{2} \left( m_{0}^{2} + m_{1}^{2} \right) \left( dL \right)^{2},
\end{eqnarray}
which lead to
\begin{equation} \begin{split}
\frac{1}{2} \braket{ \mathrm{tr}\,[ r_{\rm L} \left( dL \right) r_{\rm L}^{\dag} \left( dL \right) ]}
= \frac{1}{2} \braket{ \mathrm{tr}\,[ r_{\rm R} \left( dL \right) r_{\rm R}^{\dag} \left( dL \right) ]}
= \frac{dL}{\ell},\\
\frac{1}{2} \braket{ \mathrm{tr}\,[ t_{\rm R} \left( dL \right) t_{\rm R}^{\dag} \left( dL \right) ]}
= \frac{1}{2} \braket{ \mathrm{tr}\,[ t_{\rm L} \left( dL \right) t_{\rm L}^{\dag} \left( dL \right) ]}
= 1 - \frac{dL}{\ell}.
\end{split} \end{equation}
In contrast to classes A and AII, non-Hermiticity $\gamma_{33}$ disappears in these equations. Nevertheless, it leads to non-Hermitian delocalization with the bidirectional nature instead of the unidirectional one. To see this bidirectional delocalization, we perform the polar decomposition
\begin{equation}
t_{\rm R} = U_{\rm R} \left( \begin{array}{@{\,}cc@{\,}} 
	\sqrt{T_{+}} & 0 \\ 0 & \sqrt{T_{-}} \\
	\end{array} \right)
\end{equation}
with a unitary matrix $U_{\rm R}$, and consider the incremental changes of the transmission eigenvalues $T_{+}$ and $T_{-}$. Here, because of reciprocity, the other transmission matrix $t_{\rm L}$ is 
\begin{equation}
t_{\rm L} = \sigma_{2} t_{\rm R} \sigma_{2}^{-1}
= \left( \sigma_{2} U_{\rm R} \sigma_{2}^{-1} \right) \left( \begin{array}{@{\,}cc@{\,}} 
	\sqrt{T_{-}} & 0 \\ 0 & \sqrt{T_{+}} \\
	\end{array} \right),
\end{equation}
and hence the transmission eigenvalues are identical. Then, noticing
\begin{equation}
U_{\rm R}^{\dag} \left( L \right) \left[ t_{\rm R} \left( L+dL \right) t_{\rm R}^{\dag} \left( L+dL \right) \right] U_{\rm R} \left( L \right)
= \left( \begin{array}{@{\,}cc@{\,}} 
	\sqrt{T_{+} \left( L \right)} & 0 \\ 0 & \sqrt{T_{-} \left( L \right)} \\
	\end{array} \right) \frac{t_{\rm R} \left( dL \right) t_{\rm R}^{\dag} \left( dL \right)}{\left| 1 - r_{\rm R} \left( dL \right) r_{\rm L} \left( L \right) \right|^{2}} \left( \begin{array}{@{\,}cc@{\,}} 
	\sqrt{T_{+} \left( L \right)} & 0 \\ 0 & \sqrt{T_{-} \left( L \right)} \\
	\end{array} \right),
\end{equation}
we have the scaling equations
\begin{equation}
\frac{dT_{\pm}}{T_{\pm}} = \left( 2 m_{1} \sqrt{R_{\rm L}} \sin \varphi_{\rm L} \pm \gamma_{33} \right) dL + m_{1}^{2} \left[ \left( 4 \sin^{2} \varphi_{\rm L} - 1 \right) R_{\rm L} - 1\right] \left( dL \right)^{2}.
\end{equation}
Therefore, non-Hermiticity $\gamma_{33}$ amplifies one of the transmission eigenvalues and attenuates the other. For $L \gg \ell$, the conductances are given as
\begin{equation}
G^{\rm av}_{\rm R}/G_{\rm c} = G^{\rm av}_{\rm L}/G_{\rm c} \sim e^{\left( \left| \gamma_{33} \right| - 1/4\ell \right) L},\quad
G^{\rm typ}_{\rm R}/G_{\rm c} = G^{\rm typ}_{\rm L}/G_{\rm c} = e^{\left( \left| \gamma_{33} \right| - 1/\ell \right) L}.
\end{equation}
Consequently, eigenstates are bidirectionally delocalized for sufficiently strong non-Hermiticity $\gamma_{33}$ in contrast to both classes A and AII. This bidirectional delocalization originates from the Kramers-pair structure in class $\text{AII}^{\dag}$: when one eigenstate of a Kramers pair is delocalized toward one direction, the other is delocalized toward the opposite direction. This is sharply contrasted with the Kramers-pair structure in class AII, in which both eigenstates of a Kramers pair are delocalized toward the same direction.

\subsection{Classes AIII and $\text{AIII}^{\dag}$}

Chiral or sublattice symmetry enables the delocalization of zero modes even in Hermitian systems, accompanied by Dyson's singularity~\cite{Dyson-53}. This delocalization results from the constraint $S^{\dag} \left( E \right) = S \left( -E \right)$ on unitary scattering matrices $S \left( E \right)$ due to chiral or sublattice symmetry. In fact, $S^{\dag} \left( 0 \right) = S \left( 0 \right)$ is respected for the zero modes and the reflection matrices become Hermitian (i.e., $r_{\rm L}^{\dag} = r_{\rm L}$, $r_{\rm R}^{\dag} = r_{\rm R}$). Then, the phases of the reflection matrices are confined to be $0$ or $\pi$, which contrasts with the standard classes. Consequently, the zero modes are delocalized even in one dimension and the conductances for $L \gg \ell$ are given as~\cite{Brouwer-98, *Brouwer-00}
\begin{equation}
G^{\rm av}/G_{\rm c} \sim \sqrt{2\ell/\pi L},\quad
G^{\rm typ}/G_{\rm c} \sim e^{- \sqrt{8L/\pi\ell}}.
	\label{Seq: conductances - chiral}
\end{equation}
In the following, we consider the influence of non-Hermiticity on the delocalization due to chiral or sublattice symmetry.

We investigate a non-Hermitian continuum model
\begin{equation}
h = -\ii \tau_{3} \partial_{x} + \ii \left( \gamma_{1}/2 \right) \tau_{1} + m_{2} \left( x \right) \tau_{2},
\end{equation}
which respects chiral symmetry $\tau_{1} h^{\dag} \tau_{1}^{-1} = - H$. Notably, the non-Hermitian term $\ii \left( \gamma_{3}/2 \right) \tau_{3}$ in Eq.~(\ref{Seq: Hamiltonian - A}) is not allowed because of chiral symmetry. The nonunitary scattering matrix $S_{dL} \left( E \right)$ of a thin slice of this system is given as
\begin{equation}
S_{dL} \left( E \right)
:= \left( \begin{array}{@{\,}cc@{\,}} 
	r_{\rm L} \left( dL \right) & t_{\rm L} \left( dL \right) \\ t_{\rm R} \left( dL \right) & r_{\rm R} \left( dL \right) \\
	\end{array} \right)
= \left( \begin{array}{@{\,}cc@{\,}} 
	\left( m_{2} + \gamma_{1}/2 \right) dL & 1 +\ii E\,dL - m_{2}^{2} \left( dL \right)^{2}/2 \\ 1 +\ii E\,dL - m_{2}^{2} \left( dL \right)^{2}/2 & - \left( m_{2} - \gamma_{1}/2 \right) dL \\
	\end{array} \right).
\end{equation}
Despite $S_{dL}^{\dag} \left( E \right) S_{dL} \left( E \right) \neq 1$, this nonunitary scattering matrix is indeed Hermitian for zero modes (i.e., $E=0$). The incremental change of the transmission amplitude $T_{\rm R}$ for these zero modes is 
\begin{equation}
\frac{dT_{\rm R}}{T_{\rm R}} \simeq 
- 2 \left( m_{2} - \gamma_{1}/2 \right) \sqrt{R_{\rm L}} \left( \cos \varphi_{\rm L} \right) dL
+ m_{2}^{2} \left[ \left( 4 \cos^{2} \varphi_{\rm L} - 1\right) R_{\rm L} - 1\right] \left( dL \right)^{2}.
\end{equation}
Since the phase $\varphi_{\rm L}$ of $r_{\rm L}$ is assumed to be independently and uniformly distributed over $\left\{ 0, \pi \right\}$, we have $\braket{\cos \phi_{\rm L}} = \braket{\sin \phi_{\rm L}} = 0$, $\braket{\cos^{2} \phi_{\rm L}} = \braket{\sin^{2} \phi_{\rm L}} = 1$, and hence 
\begin{equation}
\frac{\braket{dT_{\rm R}}}{dL}
= - \frac{T_{\rm R} \left( 1 - 3 R_{\rm L}\right)}{\ell},\quad
\frac{\braket{\left( dT_{\rm R} \right)^{2}}}{dL}
= \frac{4T_{\rm R}^{2} R_{\rm L}}{\ell}.
\end{equation}
Similarly, the incremental change of the moments of the reflection amplitude $R_{\rm L}$ is
\begin{equation}
\frac{\braket{dR_{\rm L}}}{dL}
= \frac{\left( 1 - R_{\rm L}\right) \left( 1 - 3 R_{\rm L}\right)}{\ell},\quad
\frac{\braket{\left( dR_{\rm L} \right)^{2}}}{dL}
= \frac{4R_{\rm L} \left( 1 - R_{\rm L}\right)^{2}}{\ell}.
\end{equation}
In these scaling equations, non-Hermiticity $\gamma_{1}$ is not relevant to the transmission or the reflection amplitude. Thus, the conductances for the zero modes are given by Eq.~(\ref{Seq: conductances - chiral}), and the universality of non-Hermitian localization in class AIII is the same as the Hermitian counterpart.

On the other hand, non-Hermiticity changes the universality of Anderson localization in class $\text{AIII}^{\dag}$. We investigate a non-Hermitian continuum model 
\begin{equation}
h = \left( -\ii \partial_{x} + \ii \gamma_{3}/2 \right) \tau_{3} + m_{2} \left( x \right) \tau_{2},
\end{equation}
which respects sublattice symmetry $\tau_{1} h \tau_{1}^{-1} = - H$. In contrast to chiral symmetry, the non-Hermitian term  $\ii \left( \gamma_{3}/2 \right) \tau_{3}$ is allowed even in the presence of sublattice symmetry. Consequently, the unidirectional delocalization is possible in a similar manner to class A, and the conductances are given as
\begin{equation}
G^{\rm av}/G_{\rm c} \sim \sqrt{2\ell/\pi L}\,e^{\pm \gamma_{3}L},\quad
G^{\rm typ}/G_{\rm c} \sim e^{\pm \gamma_{3} L - \sqrt{8L/\pi\ell}}.
\end{equation}

It is also notable that the non-Hermitian delocalization shares the same nature with the delocalization of zero modes due to chiral symmetry. To see this correspondence, we notice the following Hermitian Hamiltonian $\tilde{H} \left( E \right)$ constructed from a non-Hermitian Hamiltonian $H$ and $E \in \mathbb{C}$~\cite{Feinberg-97, Brouwer-98, *Brouwer-00, Gong-18, OKSS-20}:
\begin{equation}
\tilde{H} \left( E \right) := \left( \begin{array}{@{\,}cc@{\,}} 
	0 & H-E \\ H^{\dag}-E^{*} & 0 \\
	\end{array} \right).
\end{equation}
When $E$ is an eigenenergy of $H$ and $\ket{\psi}$ is the corresponding right eigenstate, $\left( 0~\ket{\psi} \right)^{T}$ is a zero mode of $\tilde{H} \left( E \right)$. Thus, delocalized eigenstates can appear in $H$ even in the presence of disorder if the corresponding zero modes are delocalized in $\tilde{H} \left( E \right)$. Consistently, when $H$ belongs to class A or $\text{AII}^{\dag}$ ($\text{AI}^{\dag}$), $\tilde{H} \left( E \right)$ belongs to class AIII or DIII (CI)~\cite{KSUS-19}, in which delocalization of zero modes is possible (impossible)~\cite{Brouwer-98, *Brouwer-00}.

\section{SIV.~Non-Hermitian localization on lattices}

We investigate localization of non-Hermitian disordered systems on lattices. In particular, we consider one-dimensional Hamiltonians with onsite disorder described by
\begin{equation}
\hat{H} = \sum_{n} \left[ - \frac{1}{2} \left( \hat{c}_{n+1}^{\dag} \hop_{\rm R} \hat{c}_{n} + \hat{c}_{n}^{\dag} \hop_{\rm L} \hat{c}_{n+1} \right) + \hat{c}_{n}^{\dag} M_{n} \hat{c}_{n} \right],
\end{equation}
where $\hat{c}_{n}$ ($\hat{c}_{n}^{\dag}$) annihilates (creates) an $N$-component fermion at site $n$, and $\hop_{\rm R}$ ($\hop_{\rm L}$) and $M_{n}$ are $N \times N$ matrices that describe the hopping from the left to the right (from the right to the left) and the disordered potential at site $n$, respectively. In the presence of Hermiticity $\hat{H}^{\dag} = \hat{H}$, we have $\hop_{\rm R}^{\dag} = \hop_{\rm L}$ and $M_{n}^{\dag} = M_{n}$. When disorder is sufficiently strong, eigenstates are localized. The site-$n$ component of an eigenstate localized around $n=n_{0}$ is generally described by
\begin{equation}
\psi_{n} \sim \begin{cases}
e^{- \left| n-n_{0}\right|/\xi_{\rm L}} & \left( n < n_{0} \right), \\
e^{- \left| n-n_{0}\right|/\xi_{\rm R}} & \left( n > n_{0} \right). \\
\end{cases}
\end{equation}

The localization lengths $\xi_{\rm L}$ and $\xi_{\rm R}$ can be efficiently obtained by the transfer-matrix method~\cite{Ohtsuki-review}. We begin with the Schr\"odinger equation
\begin{equation}
- \frac{\hop_{\rm R}}{2} \psi_{n-1} - \frac{\hop_{\rm L}}{2} \psi_{n+1} + M_{n} \psi_{n} = E\,\psi_{n},
\end{equation}
where $E \in \mathbb{C}$ is an eigenenergy and $\psi_{n}$ is the site-$n$ component of the corresponding eigenstate. This leads to 
\begin{equation}
\left( \begin{array}{@{\,}c@{\,}} 
	\psi_{n+1} \\ \psi_{n} \\
	\end{array} \right) = M_{{\rm L}n} \left( \begin{array}{@{\,}c@{\,}} 
	\psi_{n} \\ \psi_{n-1} \\
	\end{array} \right),\quad
M_{{\rm L}n} := \left( \begin{array}{@{\,}cc@{\,}} 
	-2 \hop_{\rm L}^{-1} \left( E - M_{n} \right) & - \hop_{\rm L}^{-1} \hop_{\rm R} \\
	1 & 0 \\
	\end{array} \right).
\end{equation}
Then, the left localization length $\xi_{\rm L}$ is given as the inverse of the smallest positive eigenvalue of the $2N \times 2N$ matrix
\begin{equation}
\frac{1}{2L} \log \left[ \left( \prod_{n=1}^{L} M_{{\rm L}n} \right) \left( \prod_{n=1}^{L} M_{{\rm L}n} \right)^{\dag} \right].
\end{equation}
Similarly, we have
\begin{equation}
\left( \begin{array}{@{\,}c@{\,}} 
	\psi_{n-1} \\ \psi_{n} \\
	\end{array} \right) = M_{{\rm R}n} \left( \begin{array}{@{\,}c@{\,}} 
	\psi_{n} \\ \psi_{n+1} \\
	\end{array} \right),\quad
M_{{\rm R}n} := \left( \begin{array}{@{\,}cc@{\,}} 
	-2 \hop_{\rm R}^{-1} \left( E - M_{n} \right) & - \hop_{\rm R}^{-1} \hop_{\rm L} \\
	1 & 0 \\
	\end{array} \right),
\end{equation}
from which we can obtain the right localization length $\xi_{\rm R}$. 

While we have $\hop_{\rm R}^{\dag} = \hop_{\rm L}$ and hence $\xi_{\rm L} = \xi_{\rm R}$ in Hermitian systems, we can have two different localization lengths (i.e., $\xi_{\rm L} \neq \xi_{\rm R}$) in non-Hermitian systems. As discussed in the main text, the distinction between $\xi_{\rm L}$ and $\xi_{\rm R}$ leads to the two-parameter scaling of non-Hermitian localization. Moreover, we have $\left| \det M_{{\rm L}n} \right| = \left| \det M_{{\rm R}n} \right| = 1$ in Hermitian systems, which ensures $\xi_{\rm L} = \xi_{\rm R} < \infty$ and hence the absence of delocalization in one dimension. However, this is not the case in non-Hermitian systems, which enables the divergence of the localization length and the consequent delocalization even in one dimension.

\subsection{Hatano-Nelson model (class A)}

\begin{figure}[b]
\centering
\includegraphics[width=144mm]{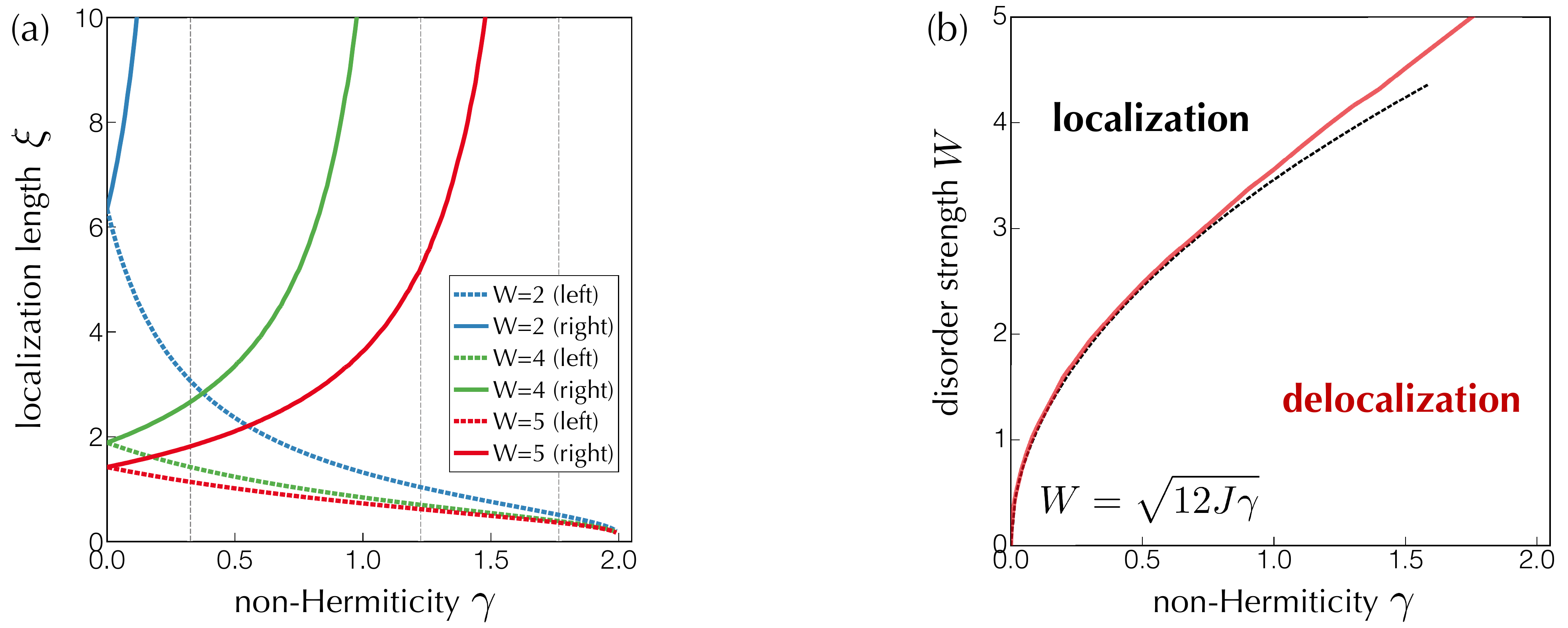} 
\caption{Localization transition in the Hatano-Nelson model ($L=5000, \hop=1.0, E=0$). Each datum is averaged over $1000$ samples. (a)~Localization length. For $\gamma \geq 0$, the right localization length diverges at a transition point $\gamma = \gamma_{\rm c}$, whereas the left localization length remains finite. The transition points (dotted lines) are $\gamma_{\rm c} = 0.325$ ($W = 2.0$), $\gamma_{\rm c} = 1.23$ ($W = 4.0$), and $\gamma_{\rm c} = 1.77$ ($W = 5.0$). (b)~Phase diagram. The red solid curve shows the numerically obtained phase boundary. For sufficiently small $\gamma$ and $W$, the phase boundary is given as $W = \sqrt{12J\gamma}$ (black dotted curve).}
	\label{Sfig: HN-supplement}
\end{figure}

We investigate the Hatano-Nelson model ($\hop_{\rm R} = \hop + \gamma/2$, $\hop_{\rm L} = \hop - \gamma/2$, $M_{n} = m_{n}$)~\cite{Hatano-Nelson-96, *Hatano-Nelson-97, *Hatano-Nelson-98}
\begin{equation}
\hat{H} = \sum_{n} \left\{ - \frac{1}{2} \left[ \left( \hop + \frac{\gamma}{2} \right) \hat{c}_{n+1}^{\dag} \hat{c}_{n} + \left( \hop - \frac{\gamma}{2} \right) \hat{c}_{n}^{\dag} \hat{c}_{n+1} \right] + m_{n} \hat{c}_{n}^{\dag} \hat{c}_{n} \right\}
\quad \left( \hop, \gamma, m_{n} \in \mathbb{R} \right).
\end{equation}
Here, the disordered potential $m_{n}$ is uniformly distributed over $\left[ -W/2, W/2 \right]$ with $W \geq 0$. The localization lengths as a function of the disorder strength $W$ are shown in Fig.~2\,(a) in the main text, and those as a function of non-Hermiticity $\gamma$ are shown in Fig.~\ref{Sfig: HN-supplement}\,(a). Fog $\gamma \geq 0$, the right localization length $\xi_{\rm R}$ diverges at a critical point $W = W_{\rm c}$ or $\gamma = \gamma_{\rm c}$, whereas the left localization length $\xi_{\rm L}$ remains finite. This is a signature of the unidirectional delocalization and is consistent with the two-parameter scaling theory of conductances for continuum models. The phase diagram is shown in Fig.~\ref{Sfig: HN-supplement}\,(b).

The nature of the unidirectional delocalization is understood by the $\text{GL} \left(1\right)$-gauge transformation (imaginary-gauge transformation in Ref.~\cite{Hatano-Nelson-96, *Hatano-Nelson-97, *Hatano-Nelson-98}). With the new fermion operators by the $\text{GL} \left(1\right)$-gauge transformation
\begin{equation}
\hat{f}_{n} := e^{-n\theta} \hat{c}_{n},\quad
\hat{f}_{n}^{\dag} := e^{n\theta} \hat{c}_{n}^{\dag}\quad
\left( \theta \in \mathbb{C} \right),
\end{equation}
the Hamiltonian reads
\begin{equation}
\hat{H} = \sum_{n} \left\{ - \frac{1}{2} \left[ \left( \hop + \frac{\gamma}{2} \right) e^{-\theta} \hat{f}_{n+1}^{\dag} \hat{f}_{n} + \left( \hop - \frac{\gamma}{2} \right) e^{\theta} \hat{f}_{n}^{\dag} \hat{f}_{n+1} \right] + m_{n} \hat{f}_{n}^{\dag} \hat{f}_{n} \right\}.
\end{equation}
Here, choosing $\theta$ such that
\begin{equation}
\left( \hop + \frac{\gamma}{2} \right) e^{-\theta} = \left( \hop - \frac{\gamma}{2} \right) e^{\theta} = \tilde{\hop},
\end{equation}
i.e.,
\begin{equation}
\theta = \frac{1}{2} \log \left( \frac{\hop + \gamma/2}{\hop - \gamma/2} \right),\quad
\tilde{\hop} = \sqrt{\hop^{2} - \left( \gamma/2\right)^{2}},
\end{equation}
we have the Hermitian Anderson model
\begin{equation}
\hat{H} = \sum_{n} \left[ - \frac{\tilde{\hop}}{2} \left( \hat{f}_{n+1}^{\dag} \hat{f}_{n} + \hat{f}_{n}^{\dag} \hat{f}_{n+1} \right) + m_{n} \hat{f}_{n}^{\dag} \hat{f}_{n} \right].
	\label{Seq: Hermitian Anderson - A}
\end{equation}
Thus, the localization lengths of the Hatano-Nelson model are
\begin{equation}
\xi_{\rm L} = \left( \xi_{0}^{-1} + \theta \right)^{-1},\quad
\xi_{\rm R} = \left( \xi_{0}^{-1} - \theta \right)^{-1},
\end{equation}
where $\xi_{0}$ is the localization length of the Hermitian Anderson model in Eq.~(\ref{Seq: Hermitian Anderson - A}). For $\gamma \geq 0$ ($\gamma \leq 0$), the localization length $\xi_{\rm R}$ ($\xi_{\rm L}$) diverges at $\gamma = \gamma_{\rm c}$ such that $\xi_{0}^{-1} = \left| \theta \left( \gamma_{\rm c} \right) \right|$. Around this critical point, we have
\begin{equation}
\xi \sim \frac{1}{\left| \theta' \left( \gamma_{\rm c} \right) \left( \gamma - \gamma_{\rm c} \right) \right|} \propto \left| \gamma - \gamma_{\rm c} \right|^{-1}.
\end{equation}
For sufficiently weak disorder, we have
\begin{equation}
\xi_{0}^{-1} \simeq \frac{\braket{m_{n}^2}}{2\,( \tilde{J}^2 - E^2 )}.
\end{equation}
When $m_{n}$ is uniformly distributed over $\left[ -W/2, W/2 \right]$, we have $\braket{m_{n}^2} = W^2/12$. Then, the critical point is given as
\begin{equation}
\left| \gamma_{\rm c} \right| \simeq \frac{W^2}{12J}
\end{equation}
for the band center $E = 0$. This is consistent with the result for the continuum model, as well as the numerical result in Fig.~\ref{Sfig: HN-supplement}\,(b).

\subsection{Non-Hermitian Anderson model with random gain or loss (class $\text{AI}^{\dag}$)}

Non-Hermitian Hamiltonians $\hat{H}$ in class $\text{AI}^{\dag}$ (orthogonal class) respect reciprocity defined by $\hat{\mathcal{T}} \hat{H} \hat{\mathcal{T}}^{-1} = \hat{H}^{\dag}$ with an antiunitary operator $\hat{\mathcal{T}}$. When $\hat{\mathcal{T}}$ is complex conjugation, reciprocity means $\hop_{\rm R}^{T} = \hop_{\rm L}$ and $M_{n}^{T} = M_{n}$. Consequently, we have $\xi_{\rm L} = \xi_{\rm R}$ and $\left| \det M_{\rm{L}n} \right| = \left| \det M_{\rm{R}n} \right| = 1$, which imposes $\xi_{\rm L} = \xi_{\rm R} < \infty$ and forbids delocalization even in the presence of non-Hermiticity.

\begin{figure}[t]
\centering
\includegraphics[width=144mm]{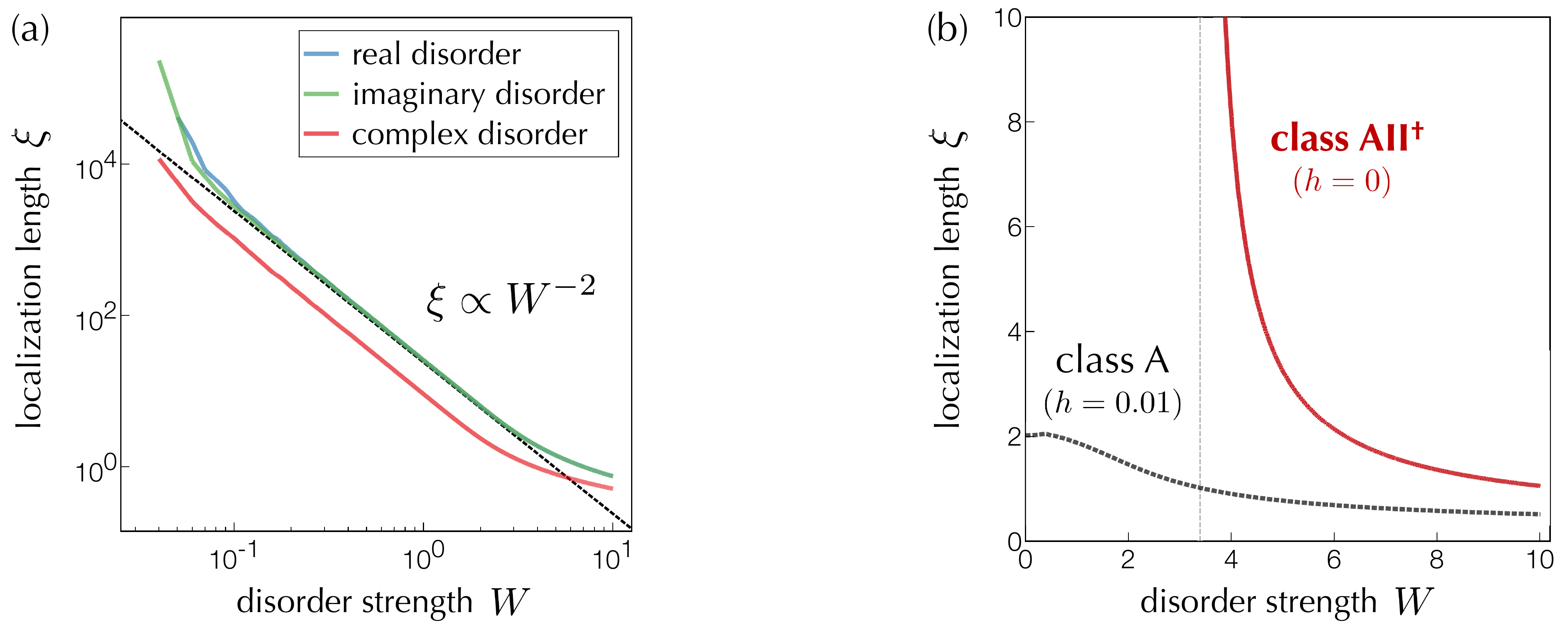} 
\caption{Localization lengths in non-Hermitian disordered systems on one-dimensional lattices with reciprocity ($L=5000, \hop=1.0, E=0$). Each datum is averaged over $1000$ samples. (a)~Non-Hermitian Anderson model with random gain or loss (class $\text{AI}^{\dag}$). For all the types of Hermitian and non-Hermitian disorder, no transition occurs and the universality class is the same. The dotted line shows $\xi = 24J^2/W^{2}$. (b)~Symplectic Hatano-Nelson model (class $\text{AII}^{\dag}$; $\gamma = 1.0, \Delta = 0.1$). In contrast to the Hatano-Nelson model without symmetry protection, both right and left localization lengths diverge at a transition point $W_{\rm c} = 3.39$ (red solid curve). Because of the reciprocity-protected nature of the delocalization, even a small reciprocity-breaking perturbation $h = 0.01$ vanishes the delocalization (black dotted curve).}
	\label{Sfig: reciprocity-supplement}
\end{figure}

In particular, we investigate the non-Hermitian Anderson model with random gain or loss ($\hop_{\rm R} = \hop_{\rm L} = \hop$, $M_{n} = m_{n} + \ii \gamma_{n}$)
\begin{equation}
\hat{H} = \sum_{n} \left\{ - \frac{\hop}{2} \left( \hat{c}_{n+1}^{\dag} \hat{c}_{n} + \hat{c}_{n}^{\dag} \hat{c}_{n+1} \right) + \left( m_{n} + \ii \gamma_{n} \right) \hat{c}_{n}^{\dag} \hat{c}_{n} \right\}\quad \left( \hop, m_{n}, \gamma_{n} \in \mathbb{R}\right).
\end{equation}
We here consider the following three types of disorder:
\begin{itemize}
\item real disorder ($m_{n} \in \left[ - W/2, W/2\right]$, $\gamma_{n} = 0$)
\item imaginary disorder ($m_{n} = 0$, $\gamma_{n} \in \left[ - W/2, W/2\right]$)
\item complex disorder ($m_{n}, \gamma_{n} \in \left[ - W/2, W/2\right]$)
\end{itemize}
Figure~\ref{Sfig: reciprocity-supplement}\,(a) shows the localization length $\xi$ as a function of the disorder strength $W$. For all the types of Hermitian and non-Hermitian disorder, $\xi$ remains finite even for small $W$, and no delocalization occurs. Moreover, $\xi$ gets smaller in proportion to $W^{-2}$ in the same manner as the Hermitian case. These results are consistent with the scaling theory of conductances for continuum models, which demonstrates that the universality of the non-Hermitian localization transitions in class $\text{AI}^{\dag}$ is the same as the Hermitian counterpart.

\subsection{Symplectic Hatano-Nelson model (class $\text{AII}^{\dag}$)}

We investigate a symplectic (reciprocal) generalization of the Hatano-Nelson model ($\hop_{\rm R} = \hop + \gamma \sigma_{3}/2 - \ii \Delta \sigma_{1}$, $\hop_{\rm L} = \hop - \gamma \sigma_{3}/2 + \ii \Delta \sigma_{1}$, $M_{n} = m_{n} + h\sigma_{3}$)
\begin{equation}
\hat{H} = \sum_{n} \left\{ - \frac{1}{2} \left[ \hat{c}_{n+1}^{\dag} \left( \hop + \frac{\gamma\sigma_{3}}{2} - \ii \Delta \sigma_{1} \right) \hat{c}_{n} + \hat{c}_{n}^{\dag} \left( \hop - \frac{\gamma\sigma_{3}}{2} + \ii \Delta \sigma_{1} \right) \hat{c}_{n+1} \right] + \hat{c}_{n}^{\dag} \left( m_{n} + h\sigma_{3} \right) \hat{c}_{n} \right\}\quad \left( \hop, \gamma, \Delta, m_{n}, h \in \mathbb{R}\right).
\end{equation}
For $h = 0$, the Hamiltonian respects reciprocity [i.e., $( \sigma_{2} \mathcal{K} )\,\hat{H}\,( \sigma_{2} \mathcal{K} )^{-1} = \hat{H}^{\dag}$ with complex conjugation $\mathcal{K}$] and hence belongs to class $\text{AII}^{\dag}$. As a result of reciprocity, the left localization length $\xi_{\rm L}$ coincides with the right localization length $\xi_{\rm R}$. On the other hand, a magnetic field $h \neq 0$ breaks reciprocity. The disordered potential $m_{n}$ is uniformly distributed over $\left[ -W/2, W/2 \right]$ with $W \geq 0$. The non-Hermitian skin effect of this model without disorder ($W = 0$) was investigated in Ref.~\cite{OKSS-20}. 

The localization lengths as a function of the non-Hermiticity $\gamma$ are shown in Fig.~2\,(b) in the main text, and those as a function of the disorder strength $W$ are shown in Fig.~\ref{Sfig: reciprocity-supplement}\,(b). In contrast to the original Hatano-Nelson model without symmetry protection, both left and right localization lengths diverge at a critical point, which is consistent with the bidirectional delocalization predicted by the scaling theory of conductances for continuum models. As a consequence of the reciprocity-protected nature of the delocalization, even a small reciprocity-breaking perturbation $h \neq 0$ vanishes the delocalization. Although $\xi_{\rm L}$ is different from $\xi_{\rm R}$ in the absence of reciprocity, no significant difference can be seen for such a small perturbation as $h=0.01$ considered in Figs.~2\,(b) and \ref{Sfig: reciprocity-supplement}\,(b).

The nature of the bidirectional delocalization is understood by the following $\text{SL} \left( 2 \right)$-gauge transformation 
\begin{equation}
\hat{f}_{n} := \left( \begin{array}{@{\,}cc@{\,}} 
	e^{-n\theta} & 0 \\ 0 & e^{n\theta} \\
	\end{array} \right) V^{-1} \hat{c}_{n},\quad
\hat{f}_{n}^{\dag} := \hat{c}_{n}^{\dag} V \left( \begin{array}{@{\,}cc@{\,}} 
	e^{n\theta} & 0 \\ 0 & e^{-n\theta} \\
	\end{array} \right)\quad
\left[ \theta \in \mathbb{C},~V \in \text{SL} \left( 2 \right) \right].
\end{equation}
With the new fermion operators $\hat{f}_{n}$ and $\hat{f}_{n}^{\dag}$, the Hamiltonian without the magnetic field (i.e., $h=0$) reads
\begin{eqnarray}
\hat{H} &=& \sum_{n} \left\{ -\frac{1}{2} \left[ \hat{f}_{n+1}^{\dag} \left( \begin{array}{@{\,}cc@{\,}} 
	e^{-\left( n+1 \right) \theta} & 0 \\ 0 & e^{\left( n+1 \right) \theta}
	\end{array} \right) V^{-1} \left( \hop+\frac{\gamma\sigma_{3}}{2} -\ii \Delta \sigma_{1} \right) V \left( \begin{array}{@{\,}cc@{\,}} 
	e^{n \theta} & 0 \\ 0 & e^{-n \theta}
	\end{array} \right) \hat{f}_{n} \right. \right. \nonumber \\ 
&&\qquad\qquad \left. \left. + \hat{f}_{n}^{\dag} \left( \begin{array}{@{\,}cc@{\,}} 
	e^{-n \theta} & 0 \\ 0 & e^{n \theta}
	\end{array} \right) V^{-1} \left( \hop-\frac{\gamma\sigma_{3}}{2} +\ii \Delta \sigma_{1} \right) V \left( \begin{array}{@{\,}cc@{\,}} 
	e^{\left( n+1 \right) \theta} & 0 \\ 0 & e^{-\left( n+1 \right) \theta}
	\end{array} \right) \hat{f}_{n+1} \right] + m_{n} \hat{f}_{n}^{\dag} \hat{f}_{n} \right\}.
\end{eqnarray}
Let us choose $V$ such that it diagonalizes $\hop + \gamma \sigma_{3}/2 - \ii \Delta \sigma_{1}$, i.e.,
\begin{equation}
V^{-1} \left( \hop+\frac{\gamma\sigma_{3}}{2} -\ii \Delta \sigma_{1} \right) V
= \left( \begin{array}{@{\,}cc@{\,}} 
	\hop + \sqrt{( \gamma/2)^{2}-\Delta^{2}} & 0 \\ 0 & \hop - \sqrt{( \gamma/2)^{2}-\Delta^{2}}
	\end{array} \right).
\end{equation} 
Then, the Hamiltonian reads
\begin{eqnarray}
\hat{H} &=& \sum_{n} \left\{ -\frac{1}{2} \left[ \hat{f}_{n+1}^{\dag} \left( \begin{array}{@{\,}cc@{\,}} 
	e^{-\theta}\,( \hop + \sqrt{( \gamma/2)^{2}-\Delta^{2}}\,) & 0 \\ 0 & e^{\theta}\,( \hop - \sqrt{( \gamma/2)^{2}-\Delta^{2}}\,)
	\end{array} \right) \hat{f}_{n} \right. \right. \nonumber \\ 
&&\qquad\qquad \left. \left. \quad+ \hat{f}_{n}^{\dag} \left( \begin{array}{@{\,}cc@{\,}} 
	e^{\theta}\,( \hop - \sqrt{( \gamma/2)^{2}-\Delta^{2}}\,) & 0 \\ 0 & e^{-\theta}\,( \hop + \sqrt{( \gamma/2)^{2}-\Delta^{2}}\,)
	\end{array} \right) \hat{f}_{n+1} \right] + m_{n} \hat{f}_{n}^{\dag} \hat{f}_{n} \right\}.
\end{eqnarray}
Furthermore, let us choose $\theta$ such that it satisfies
\begin{equation}
e^{-\theta} \left( \hop + \sqrt{ \left( \gamma/2\right)^{2} - \Delta^{2}} \right)
= e^{\theta} \left( \hop - \sqrt{ \left( \gamma/2\right)^{2} - \Delta^{2}} \right)
= \tilde{\hop},
\end{equation}
i.e., 
\begin{equation}
\theta = \frac{1}{2} \log \left( \frac{\hop + \sqrt{ \left( \gamma/2\right)^{2} - \Delta^{2}}}{\hop - \sqrt{ \left( \gamma/2\right)^{2} - \Delta^{2}}} \right) ,\quad
\tilde{\hop} = \sqrt{\hop^{2} - \left( \gamma/2 \right)^{2} + \Delta^{2}}.
	\label{Seq: theta - A2}
\end{equation}
Consequently, the Hamiltonian reduces to the Hermitian Anderson model
\begin{equation}
\hat{H} = \sum_{n} \left\{ -\frac{\tilde{\hop}}{2} \left( \hat{f}_{n+1}^{\dag} \hat{f}_{n} + \hat{f}_{n}^{\dag} \hat{f}_{n+1} \right) + m_{n} \hat{f}_{n}^{\dag} \hat{f}_{n} \right\}.
	\label{Seq: Hermitian Anderson - A2}
\end{equation}
Thus, the localization length of the reciprocal Hatano-Nelson model is given as
\begin{equation}
\xi_{\rm L} = \xi_{\rm R} 
= \left( \xi_{0}^{-1} - \left| \mathrm{Re} \left( \theta \right) \right| \right)^{-1},
\end{equation}
where $\xi_{0}$ is the localization length of the Hermitian Anderson model in Eq.~(\ref{Seq: Hermitian Anderson - A2}). As seen from Eq.~(\ref{Seq: theta - A2}), $\mathrm{Re} \left( \theta \right)$ is zero for $\left| \gamma \right| \leq 2 \left| \Delta \right|$, which leads to the plateau of the localization length in Fig.~2\,(b) in the main text. Importantly, the above $\text{SL} \left( 2 \right)$-gauge transformation is unfeasible in the presence of reciprocity-breaking perturbations, which forbids the bidirectional delocalization.


\end{document}